%% file: chiral_t.tex
\documentclass[aps,physrev,reprint,amsmath,amssymb]{revtex4-2}

\usepackage{graphicx}
\usepackage{dcolumn}
\usepackage{bm}
\usepackage{hyperref}

\usepackage{docmute}

\begin{document}

\title{Chirality transitions in a system of active flat spinners}

\author{Miguel A. L\'opez-Casta\~no}
\affiliation{Departamento de F\'{\i}sica, Universidad de
  Extremadura, 06071 Badajoz, Spain}

\author{Alejandro M\'arquez Seco}
\affiliation{Departamento de
  F\'{\i}sica, Universidad de Extremadura, 06071 Badajoz, Spain}

\author{Alicia M\'arquez Seco} \affiliation{Departamento de F\'{\i}sica, Universidad de Extremadura,
  06071 Badajoz, Spain}

\author{\'Alvaro Rodr\'iguez-Rivas}
\affiliation{Department of Physical, Chemical and Natural
  Systems, Pablo de Olavide University, 41013, Sevilla, Spain}

\author{Francisco Vega Reyes}
\affiliation{Departamento de F\'{\i}sica and Instituto de
  Computaci\'on Cient\'{\i}fica Avanzada (ICCAEx), Universidad de Extremadura, 06071 Badajoz, Spain}
\email{fvega@eaphysics.xyz}

\date{\today}

\begin{abstract}
  We study in this work the 2D dynamics of an experimental system of disk-shaped rotors, fluidized
  by turbulent upflow. Contrary to previous knowledge, our experiments show the same particle chiral
  geometry can produce flows with different chiralities. In particular, we unveil a conspicuous
  complex chiral flow, which displays multiple persistent vortexes with either sign, located
  randomly in the system. This peculiar phase mediates a continuous transition, which takes place as
  the kinetic energy input increases, from a flow with positive chirality (one vortex rotating in
  the same direction as particles spin) to a flow with negative chirality (one vortex in opposite
  sense to particle spin). We find that these surprising transitions are determined by the specific
  state of the statistical correlations between particle spin and translational velocity. We discuss
  how these correlations are determined in turn by the combined action of a series of mechanisms
  (for instance, heat dissipation at the boundaries, particle activity, average kinetic
  energy\dots), several of which are unveiled here.
\end{abstract}

\maketitle

\section{Introduction}

Chiral fluids have received much attention in recent years due to their complex and non-trivial
dynamics. Examples can be found in a variety of contexts and scales: biology \cite{WHB11}, colloids
\cite{SBSI19}, granular matter \cite{TYRGL05}, etc. These chiral fluids are all composed of many
particles with some kind of geometric or dynamic asymmetry, which breaks the system symmetry under
parity and temporal inversion, i.e., they are composed of particles whose geometrical/dynamical configuration
has \textit{chirality}. 

An important consequence of this chirality is that hydrodynamic theory for conventional fluids does
not describe the complex behavior observed in these fluids. This is so, mainly, because an
anti-symmetric component of the stress tensor (absent in regular fluids) emerges in fluids composed
of chiral particles. It is well established \cite{BSAV17} that the anti-symmetric stress is
responsible for the development of chiral flow, which in turn can allow for the emergence of
hyperuniform structures \cite{LN19,DAD18}, collective phenomena \cite{NKEG14} such as flocking or
swarms \cite{ZSS20} or topological effects \cite{SZBV17}. Therefore, understanding and control of
the parameters that govern the complex dynamics of these special fluids can help deepen the
knowledge of important biological processes and also allow for the development of applications, such
as smart materials \cite{Yetal21,SZBV17}. However, the mechanisms that, at mesoscopic level, give
rise to the emergence of specific properties of this anti-symmetric tensor (which is related to new
transport coefficients \cite{A98}) are not well understood yet, and as a consequence, the topology
of chiral flow is usually analyzed according to the geometrical configuration of particles
alone. For instance, the role of the fluidization state of the system and its statistical
correlations remains to be determined.


In order to analyze this important question, we study the statistical properties in a set of
macroscopic flat spinners, which can be regarded as a simple test system that can reproduce some
important aspects of the dynamics in more complex set-ups composed of active rotor particles
\cite{SBSI19,ZSS20,HKTGAS20,ZPIBV16,SL20}. Active spinners are relevant since they allow to study a
variety of phenomena such as chiral crystals \cite{HML20}, hexatic phase \cite{YH15}, discontinuous
chiral flow transition under density changes \cite{WRDD18}, special topological properties
\cite{BSAV17,SZBV17,DMV18,SDFVV19}, or ferromagnetically coupled rotation \cite{CTCD20}.

More specifically, and in order to describe the dynamics of chiral particles for a system that is
more akin to biological systems, we conduct here an investigation on the phase behavior of a system
of flat rotors with constant sign of their average spin (i.e., constant particle chirality sign). In
nature, as it is known, most active particles have a definite chemical/geometrical configuration;
i.e., the chirality of most active particles is well defined and does not change. And this is
actually very relevant since, for instance, many isomers that differ only in the chirality sign play
significantly different roles in the most relevant bio-chemical and bio-physical processes,
according to particle chirality alone \cite{M73}.

In several previous works, an important phenomenology of spontaneous changes in the direction of
vortex rotations (i.e., inversion of the chirality of the flow) has been observed
\cite{GAS17,KS18,BLB20,ZYY20}. However, this inversion appears to be always related to different
inversion mechanisms of the chirality of the constituent particles. In effect, inversion of the
chirality of the flow has been reported for particles with oscillating/random chirality such as
magnetic spinners \cite{GAS17} and rollers \cite{KS18}, and for Janus particles \cite{BLB20}. Another
similar but not analogous situation has been described where, at constant particle
chirality sign, concentric flow rings with alternating vorticity sign appear \cite{ZYY20} (and thus,
the system does not display actual chiral vortexes but opposing alternate stream lines). In
summary, 
according to previous knowledge, a system of particles with inherently constant chirality sign (as
ours is) would not have the ability to display chiral flow inversion. As an exception, the flow
chirality does not rely on particle chirality inversion, in a recent work \cite{WRDD18}. However,
flow chirality inversion was produced exclusively under a change in the particle density of the
system. We will see that the phenomenology is actually more complex, and a plethora of transitions
at constant density are here. We discuss this is due to the fact that transitions are controlled by
additional variables, found. Furthermore, identification of all the relevant mechanisms in the
chiral transitions will allow us to discover a new phase and also previously unknown properties of
chiral flow transitions.

Our results reveal that, contrary to what would be expected, inversion of flow chirality for a
system of particles with constant chirality sign is possible and is in general to be
expected. Furthermore, we have observed that the inversion appears as a continuous transittion that
is mediated by a complex chiral flow phase (that had remained unknown so far). This complex chiral
state is characterized by a disordered spatial distribution of vortexes of different signs and
sizes. Moreover, the transition to a flow with reverse chirality sign (i.e., flow chirality is
opposed to particle chirality) occurs when the average translational kinetic energy of the particles
is above a threshold. Furthermore, the location of this threshold is determined by structural
changes in the statistical correlations of the system, whose role has not previously been reported
in flow chirality transitions (and that here we characterize by means of a theoretical analysis,
based on the description of the cumulants of the single particle distribution function, which have
not previously reported in this context either). 


\section{Experimental set-up}
\label{setup}
\begin{figure*}[htbp]
  \centering \hspace{1.5em} \includegraphics[width=0.475\textwidth]{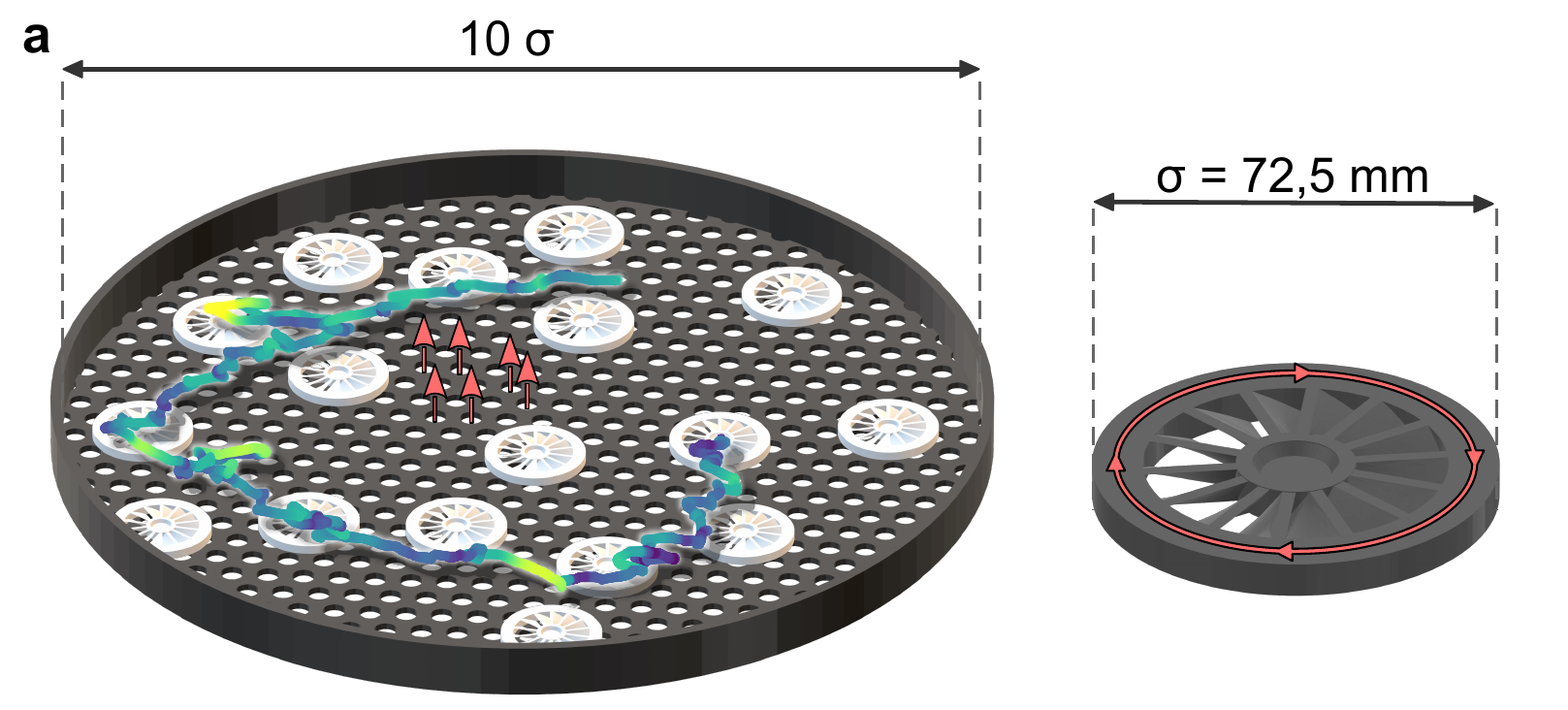}
  \hspace{1.5em} \includegraphics[width=0.275\textwidth]{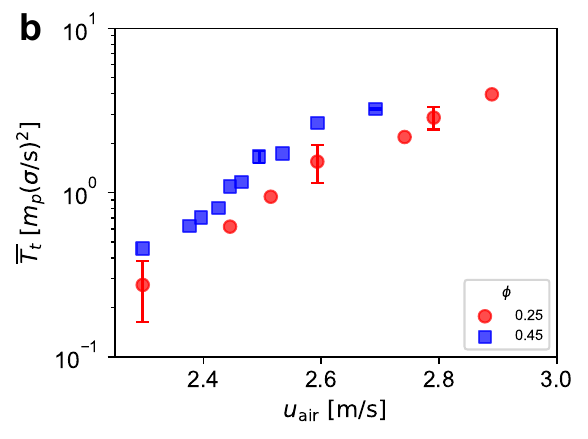} \centering
  \includegraphics[width=0.92\textwidth]{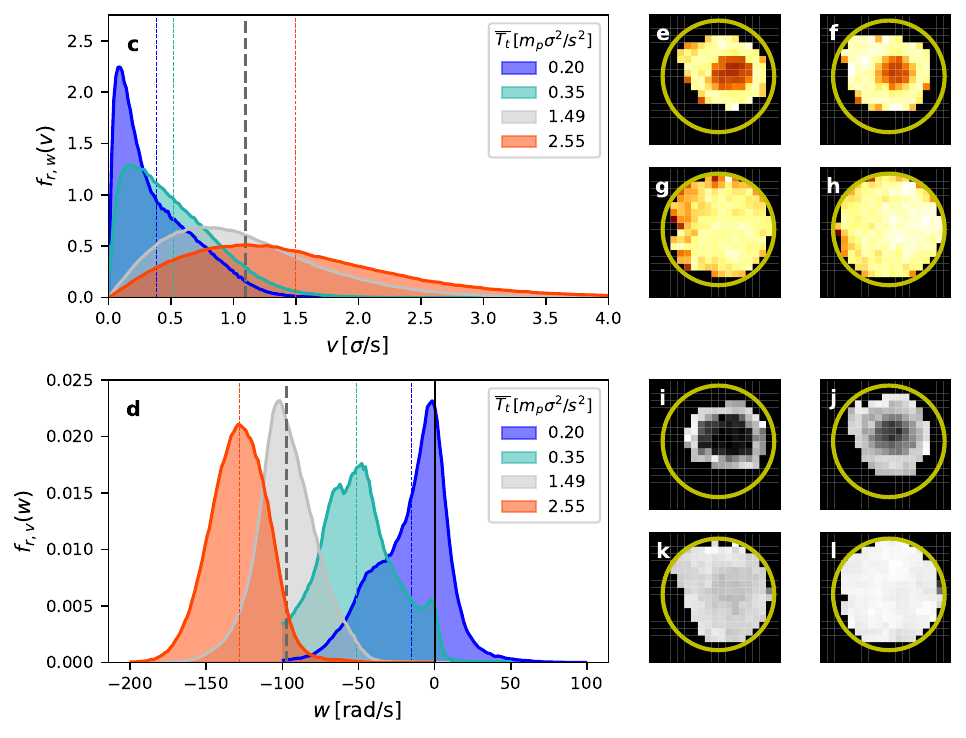}
  \caption{{(a)} {Schematic renderization of the system (left) and of one particle (right)}. The
    real trajectory of one particle is drawn in the left image, with darker/lighter color indicating
    lower/higher speed, respectively. {(b)} Global translational kinetic energy ($\overline{T_t}$)
    vs. upflow average speed ($a_\mathrm{air}$), for $\phi=0.25, 0.45$. Y axis is in log scale. As
    we can see, $\overline{T_t}\equiv\langle T_t\rangle_r$ grows exponentially with
    $u_\mathrm{air}$. {Marginal distribution functions}: {(c)} Of the particle speed ($v$),
    $f_{r,w}(v)$; {(d)} of the particle spin ($w$), $f_{r,v}(w)$. The spatial averages
    $\langle v\rangle_r$, $\overline\Omega\equiv\langle w\rangle_r$ are indicated for each
    distribution ($f_{r,w}(v)$ and $f_{r,v}(w)$, respectively) with a vertical line with the same
    color as its distribution. Also, in panel (d), a black solid line is drawn with $w=0$, in order
    to separate the reverse (counter clockwise, in this case) spin region (w>0). {Reduced 2D
      fields}, for several $T_t$ values: {(e-h)} Reduced particle speed fluctuations
    $U^*(x,y)\equiv v_\mathrm{th}(x,y)/V_0$, with $V_0=\mathrm{max}\left(v_\mathrm{th}(x,y)\right)$:
    (e) $\overline{T_t}=0.66$; (f) $\overline{T_t}=0.35$; (g) $\overline{T_t}=1.48$; (h)
    $\overline{T_t}=2.54$. {(i-l)} Reduced average spin
    $\Omega^*(x,y)\equiv\left< w(x,y)\right>/w_0$, with
    $w_0=\mathrm{max}\left(\langle w(x,y)\rangle\right)$: (i) $\overline{T_t}=0.66$; (j)
    $\overline{T_t}=0.35$; (k) $\overline{T_t}=1.48$; (l) $\overline{T_t}=2.54$.  In both $U^*(x,y)$
    and $\Omega^*(x,y)$ color maps, brighter is higher and darker is lower; black stands for
    $U^*(x,y),\Omega^*(x,y)=0$ and white for $U^*(x,y),
    \Omega^*(x,y)=1$. $\overline{T_t}\equiv\langle T_t\rangle_r$ is in units of
    $m_p\sigma^2/\mathrm{s}^2$. In (e)-(l), the system boundary is marked with a thick yellow
    line. Area fraction for panels (c)-(l): $\phi=0.25$.}
  \label{fig:distributions}
\end{figure*}

We use an experimental configuration consisting of a set of $N$ identical, flat (disk-shaped)
particles which are provided with 14 blades, as sketched in Figure~\ref{fig:distributions} (a). The
dynamics in our experiments is constrained to a horizontal perforated metallic grid. The grid is
delimited by a circular boundary of diameter $L=10~\sigma$ where $\sigma=7.25~\mathrm{cm}$ is the
particle diameter. Particle mass is $m_p=7.1~\mathrm{g}$ and
$I\simeq (1/8)~m_p\,\sigma^2=46~\mathrm{g\,cm^2}$ is the moment of inertia (approximating the true
value of $I$ to that of a homogeneous cylinder/disk). A controllable air current impinges the arena
from below.


The blades are tilted with respect to the horizontal so that a steady clockwise average spin results
from the upflow past the particles (Figure~\ref{fig:distributions}) that in turn yields vortex
shedding \cite{vD82}, thus inducing stochastic horizontal translations as well. Therefore, under
steady air upflow, the system achieves a stationary state with constant (rotational and
translational) average particle kinetic energy. 

The average kinetic energy of the particles is monotonically increasing for higher air upflow, as
Figure~\ref{fig:distributions} (b) shows. Due to particle-particle collisions and upflow turbulence,
the particle spin $\textbf{w}=w_z\mathbf{\hat e}_z$ is not exactly constant. Here,
$\mathbf{\hat e}_z$ is a unit vertical vector pointing upwards (same direction, opposite sense of
gravity), so that $w_z<0$ stands for clockwise rotation and vice versa. Due to the designed blade
tilting in the set of identical disks, $w_z<0$ most times for all particles.

As a consequence of these features, the system displays a non-equilibrium distribution function $f$,
which for an arbitrary state is $f(\mathbf{r},\mathbf{v},\mathbf{w};t)$ ($t$, is time, $\textbf{r}$
is the 2D position vector, $\textbf{v}$ is particle translational velocity) whose standard
deviations depend on air current intensity.

Moreover, as a result of the interplay of particle collisions and air upflow and, due to the
symmetries in the experimental set-up, the steady base states (i.e., the simplest steady states
\cite{VU09}) are statistically characterized by a particle distribution function of the form
$f(r,\mathbf{v},\mathbf{w})$. Moreover, the geometric asymmetry of the particles
(Figure~\ref{fig:distributions} \textbf{a}) yields a non-vanishing spin field of the form
$\bm\Omega(r)=\Omega(r)\,\mathbf{\hat e}_z$, which will be responsible for the emergence of
circulating flow of the form $\mathbf{u}=u(r)\,\mathbf{\hat e}_\phi$, where $\mathbf{\hat e}_\phi$
denotes the azimuthal direction.

In this way, a symmetry break arises so that the flow has \textit{chirality} \cite{CL00}.

 The marginal distribution functions, used in Figure 1 are defined as
 
\begin{equation}
  \begin{split}
    &f_{r,w}(v) = \int(1/n(r))\mathrm{d}\mathbf{r}\int\mathrm{d\mathbf{w}}f(r,\mathbf{v},\mathbf{w})\delta(|\mathbf{v}-v|). \\
    &f_{r,v}(w)=\int(1/n(r))\mathrm{d}\mathbf{r}\int\mathrm{d\mathbf{v}}f(r,\mathbf{v},\mathbf{w})\delta(w_z-w),
    \end{split}
    \label{f_marginal}
  \end{equation}  where
  $n(r)$ is the particle density field. The relevant fields, particle density $n(r)$, flow velocity
  $\mathbf{u}(r)$, particle average spin $\bm{\Omega}(r)$ are defined in Eqs.~\ref{fields} of Appendix~\ref{subsec:fdistr_prop}.

\section{Chirality transitions}
\label{results}

Since the sign of our particles chirality (spin) is constant, we could only expect, from previous
knowledge, that the chiral fluid displays steady flows with a chirality sign that mimics that of its
constituent particles, as modelled in theoretical works \cite{BSAV17} and extensively reported in
previous experiments. (We can only find one previous work where a transition in the sign is reported
for particle chirality with constant sign; the flow chirality reversal being observed as density
increases \cite{WRDD18}.) However, we have surprisingly detected that the direction of flow
circulation displays a complex behavior that is very sensitive to average kinetic energy of the
particles.
As we will see, we have additionally discovered that this complexity results from the balance
between two distinct mechanisms, namely particle spin-velocity statistical correlations and heat
dissipation at the outer walls.

\begin{figure*}[htbp]
  \centering \includegraphics[width=\textwidth]{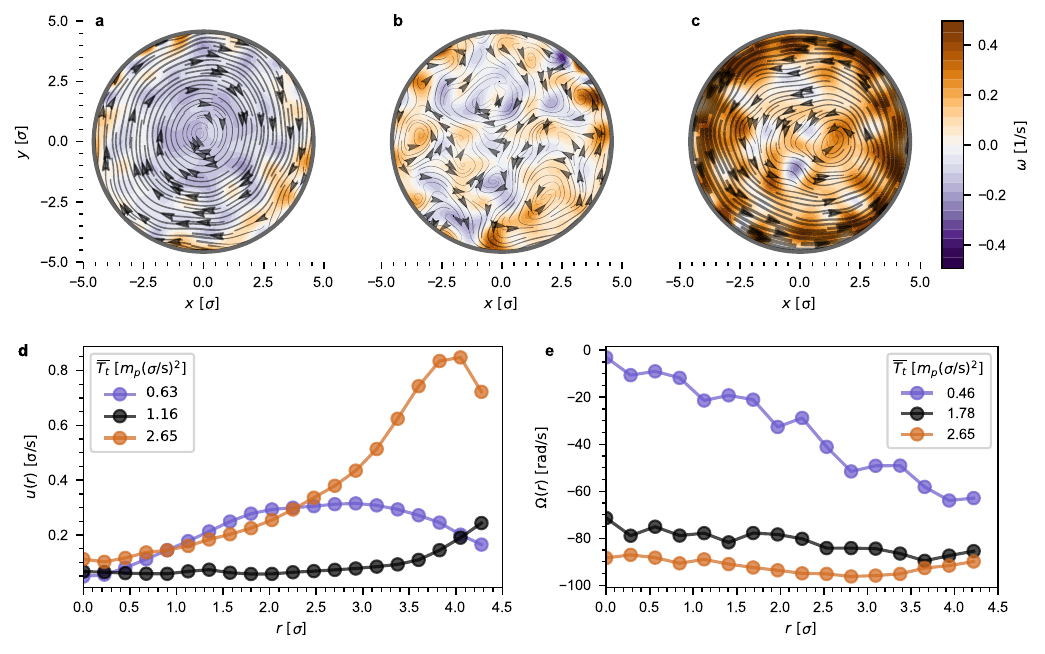}
  \caption{(a)-(c) Stream lines and vorticity field for three experiments with the same area
    fraction $\phi=0.45$ and
    $\overline{T_t}\equiv\langle T_t\rangle_r=\{ 0.63, ~1.16, ~2.65 \} ~m_p (\sigma/\mathrm{s})^{2}$
    for (a), (b), (c), respectively. They represent the different chiral flow phases we have found:
    (a) chiral flow with rotation in the same direction as the particle spin, or spinwise, (we
    denote this phase as $\mathbb{C}_+$, clockwise in this case, and thus with negative vorticity
    $\omega$), (b) complex chiral flow (note the multiple vortexes with different sense of rotation,
    we denote this phase as $\mathbb{C}_\pm$), and (c) a system with counter-spinwise fluid
    rotation, $\mathbb{C}_-$. Panel (d) shows the radial profile of the flow velocity modulus
    $u(r)$, each line corresponds to one of the experiments of the previous panels. (e) displays the
    mean particle angular velocity as we get away from the center. For spin-wise chirality
    ($\mathbb{C}_+$ ), there is a clear gradient where particles move faster near the walls and
    slower in the central region; meanwhile, in complex or counter-spin-wise situations particles
    maintain a steady self-rotation speed for every $r$, this speed being larger in magnitude the
    stronger the driving.}
  \label{fig:flow_vort}
\end{figure*}

For convenience, we use the notation
$\langle\dots\rangle= (1/n(r))\int\mathrm{d}\textbf{v}\int\mathrm{d}\textbf{w}\dots
f(r,\mathbf{v},\mathbf{w})$, where
$n(r)=\int\mathrm{d}\mathbf{v}\int\mathrm{d}\mathbf{w}f(r,\mathbf{v},\mathbf{w})$ is the particle
density. We define also the global area fraction as $\phi=N\sigma^2/L^2\simeq N/100$. We define the
field of translational kinetic energy fluctuations as $T_t(r)=(m_p/2)\langle V^2\rangle$ (where
$V^2=(\mathbf{v}-\mathbf{u})^2$, with $\mathbf{u}(r)=\langle \mathbf{v}\rangle $) and the rotational
average kinetic energy as $T_r(r)=(I/2)\langle w^2\rangle$. We define as well the spin kinetic
energy fluctuations $T_r^*(r)=(I/2)\langle W^2\rangle$ (with $W^2=(\mathbf{w}-\bm\Omega)^2$ and
$\bm{\Omega}(r)=\langle\mathbf{w}\rangle$). It will also be useful to denote spatial averaging as
$\langle\dots \rangle_r=\int\mathrm{d}\mathbf{r}\dots$.

Figures~\ref{fig:distributions}~(\textbf{c}), (\textbf{d}) show the marginal distribution functions,

\begin{equation}
  \label{marginal_distrs}
  f_{r,w}(v)=\langle \langle
  \delta(|\mathbf{v}|-v)\rangle\rangle_r,
  \quad f_{r,v}(w)=\langle\langle\delta(w_z-w)\rangle\rangle_r, 
\end{equation} whose first moment averages increase for increasing air current
(denoted as $\overline{T_t}\equiv\langle T_t\rangle_r$). In ~\ref{marginal_distrs}, $\delta(x=0)=1$ and
$\delta(x\neq0)=0$, and $|\mathbf{v}|=(v_x^2+v_y^2)^{1/2}$, and
$\mathbf{w}=w_z\mathbf{\hat e}_z$. (See Appendix~\ref{AppendixB} for formal expressions of the marginal
distributions and other relevant magnitudes.)

\begin{figure}
  \centering \includegraphics[width=0.95\columnwidth]{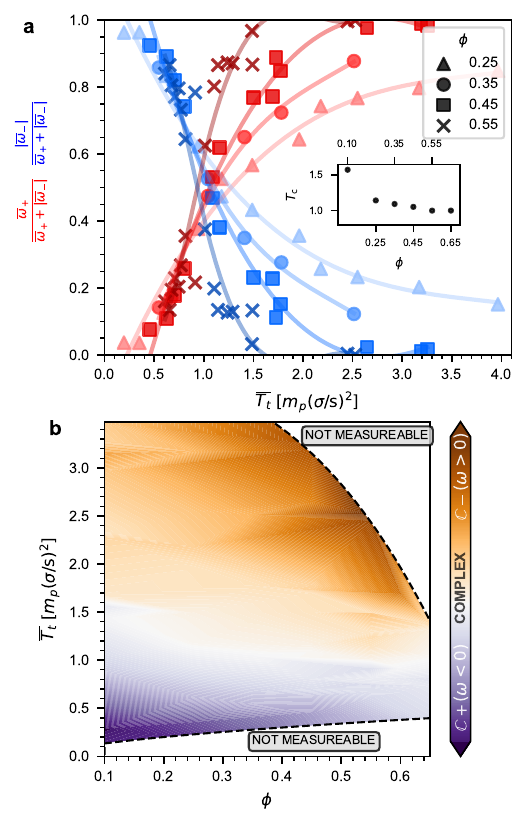}
  \caption{(a) We have represented the fraction of total positive ($\overline{\omega_{+}}$, red
    points) and negative ($| \overline{\omega_{-}} |$, blue points) vorticity for experiments at
    four representative area fractions, we added a simple spline approximating the points to show
    the trend, represented with lines. The critical average kinetic energy $T_{\mathrm{c}}$ for each
    density is found by calculating the crossing point of both fractions. The inset displays a
    slight negative slope in the value of $T_{\mathrm{c}}$ vs $\phi$, therefore we confirm a
    $\mathbb{C}+ \rightarrow \mathbb{C}-$ chirality transition at constant $T_t$, just increasing
    $\phi$ \cite{WRDD18}; nevertheless we find that $T_t$ is the main driver of changes in the
    collective rotation of spinners. (b) Chirality phase map, the color gradient indicates the
    global vorticity $\overline\omega\equiv\overline{\omega_+}+\overline{\omega_-}$ which is
    representative of the dominant rotation direction. In the purple, areas where collective motion
    goes in the same direction as the natural particle spin ($\mathbb{C}+$), experiments located in
    orange region have a counter spin wise chirality ($\mathbb{C}-$). White regions are transitional
    areas where several rolls of opposite direction are found (complex chirality, $\mathbb{C\pm}$).}
  \label{fig:phase_map}
\end{figure}

As we can see in Figures~\ref{fig:distributions}~(c), (d), the shapes of the marginal distributions
$f_{r,w}(v)$, $f_{r,v}(w)$ vary significantly versus airflow intensity. In particular, the marginal
spin distribution (Figure~\ref{fig:distributions} (d)) displays an interesting behavior. At low
activity levels (low upflow current intensity), the spin distribution extends to positive values
(counter clockwise spin). This behavior is likely due to angular momentum transfer upon particle
collisions \cite{FLCA94}, which can momentarily reverse particle spin. Arguably, the spin would
rapidly return after collision to its clockwise rotation, as imposed by the blades orientation with
respect to upflow. However, as upflow intensity increases, clockwise torque on particle blades
becomes strong enough so as to neutralize the spin reversal mechanism upon particle collisions.
Most notably, in the weak driving state, the marginal distribution displays a secondary maximum
around $w=0$. In spite of all this details in the collisional dynamics, we have consistently seen
that the average particle spin is consistently negative in all experiments; i.e., we deal with a
system for which the sign of the average particle chirality is constant, always negative (clockwise)
in this case. Therefore, any possible structural changes in the steady states cannot be due to an
eventual (non-existing) change in particle chirality. Figures~\ref{fig:distributions}~(e)-(h),
(i)-(l) the 2D density fields $v_{th}(x,y)\equiv\sqrt{\langle V^2\rangle}$ and
$\Omega(x,y)\equiv\langle w\rangle$, respectively. The (approximately) radial structure can be
seen. It is also noticeable that particles do not occupy all the available space but concentrate in
the central region. This kind of particle density behavior has not been observed at higher density
(see the corresponding Figures~\ref{fig:distributions}~(c-l) for a higher packing fraction in the
Supplementary Material \cite{suppl}). Also, it is apparent that fluidization is weaker in the center
for both particle velocity and spin, probably due to an enhanced cooling rate, due to more frequent
particle-particle collisions \cite{VSK14}.

We look up now at the fluid vorticity $\omega$. The flow velocity $\mathbf{u}$ (stream lines) and
vorticity (color map) are represented in Figures~\ref{fig:flow_vort}~(a)\textendash (c) for
different (and increasing) upflow intensities. Each panel illustrates one of the three different
vorticity behaviors we have found. It is apparent that the flow field has a broken symmetry, since
its mirror image is not identical; i.e.\ the flows observed here have \textit{chirality}
\cite{CL00}. First, in panel (a), if the system is cold enough, fluid circulates in the same
direction as the average particle spin (clockwise, in our system). We denote this behavior as
\textit{spin wise chirality} $\mathbb{C}+$). Next, at intermediate activity levels
(Figure~\ref{fig:flow_vort}~b), the system undergoes a transitional behavior with several vortexes,
each with either spin wise or counter spin wise rotation. We denote this as \textit{complex
  chirality} ($\mathbb{C}\pm$). Finally, for high driving (Figure~\ref{fig:flow_vort}~c), the system
achieves complete chirality reversal, which we denote as \textit{counter spin wise chirality}
($\mathbb{C}-$). These transitions can also be seen in Movie 1
\cite{suppl}. Figures~\ref{fig:flow_vort}~(d), (e) analyze the behavior of the flow velocity and
spin fields. In particular, for $\mathbb{C}+$ states $u(r)$ reaches its maximum at the midpoint
between the center and system boundary, whereas for $\mathbb{C}\pm$, $\mathbb{C}-$ the maximum is
reached at the system boundary. On the other hand, the spin field $\Omega(r)$ increases in magnitude
for $\mathbb{C}+$ but remains essentially constant for $\mathbb{C}\pm$ and $\mathbb{C}-$. In
summary, both fields present a distinct behavior for each of the cases. This could indicate that the
chirality mechanism is inherently different for the three chiral states
$\{\mathbb{C}+, \mathbb{C}\pm, \mathbb{C}-\}$.

Let us analyze in more detail the chirality transitions in Figure~\ref{fig:phase_map}. As we can see
in panel (a), the spatial averages of counter spin wise vorticity
$\overline{\omega_+}=\langle\omega\,\Theta(\omega)\rangle_r$ and spin wise vorticity
$\overline{\omega_-}=\langle\omega\,\Theta(-\omega)\rangle_r$ (here, $\Theta(x)=1$ for $x>0$ and
$\Theta(x)=0$ for $x<0$) present inverse behaviors; i.e., spin wise vorticity monotonically
decreases with $\overline{T_t}$ whereas counter spin wise vorticity increases monotonically. Due to
this, $\overline{\omega_+}$ and $\overline{\omega_-}$ eventually cross at a given value of
$\overline{T_t}$, that we identify as the critical point, and is slightly different for each density
(see Figure~\ref{fig:phase_map}~(a) inset). These values (denoted as $T_c$) are shown in the inset.
Partial vorticities have been reduced in Figure~\ref{fig:phase_map}~(a) with
$\overline{\omega_+} + |\overline{\omega_-}|$ and, as we can see, in this scaled representation the
curves for different densities all collapse if $T_t<T_c$ but split for $T_t>T_c$, suggesting again
that a different mechanism begins to govern the chirality behavior and that this mechanism is
related to emergent boundary currents \cite{TYRGL05}. In effect, particle-wall collision frequency
depends on the system density, which could account for this divergence.


\begin{figure*}[htbp]
  \centering \includegraphics[width=0.85\textwidth]{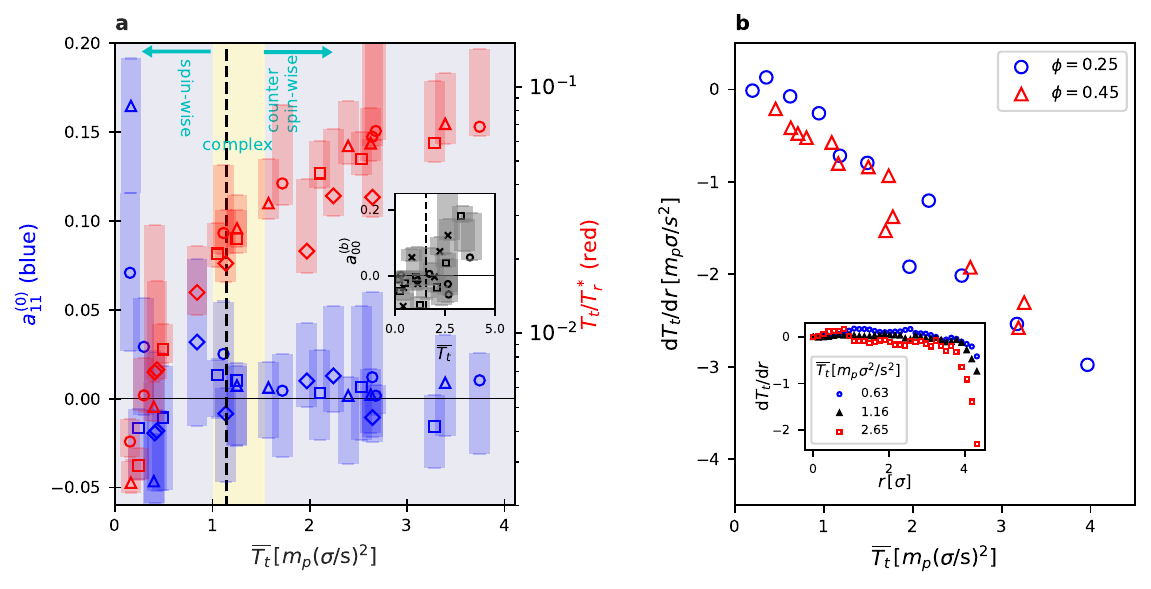}
  \caption{ {(a)} Particle velocity-spin correlations, vs.
    $\overline{T_t}\equiv\langle T_t\rangle_r$, as measured from the cumulant
    $a_{11}^{(0)}=\frac{1}{2}\left(\frac{\langle V^2W^2\rangle}{\langle V^2\rangle\langle
        W^2\rangle}-1\right)$ (blue series, left $Y$ axis), and the ratio
    ${\overline{T}_t}/{\overline{T}^*_r}$ (red series, right $Y$ axis). Inset represents cumulant
    $a_{00}^{(b)}$ (dark grey color). The error bars of series of $a_{11}^{(0)}$,
    $\overline{T}_t/\overline{T}^*_r$ and $a_{00}^{(b)}$ points are highlighted in blueish, reddish
    and grey backgrounds respectively. The transition interval in $\overline{T_t}$, with complex
    chirality, is highlighted in yellow; a dashed line marks the chirality transition point
    $T_c$. Magnitudes are measured in different annuli of the system, each annulus marked with a
    different symbol: $\circ: 0.49~\sigma $, $\triangle: 1.47~\sigma$, $\square: 2.45~\sigma$,
    $\diamond: 3.43~\sigma$ for $a_{11}^{(0)}$ and $\overline{T}_t/\overline{T}^*_r$ and
    $\circ: 0.87~\sigma $, $\square: 2.62~\sigma$, $\times: 4.36~\sigma$ for $a_{00}^{(b)}$; area
    fraction is $\phi=0.25$. {(b)} Radial gradient $\mathrm{d}T_t/\mathrm{d}r$ (heat flux
    $q_r\propto(\mathrm{d}T/\mathrm{d}r$)), as a function of $\overline{T_t}$, and measured at the
    boundary, i.e., $r\simeq 5\,\sigma$. Inset shows $\mathrm{d}T_t/\mathrm{d}r$ radial profiles for
    several values of $\overline{T_t}$. {(c)} Interaction between two spinners result in rotation
    around their center of mass, this mechanism explains the transfer of intrinsic angular momentum
    into spin-wise circulation; this happens as long as spin and velocities are statistically
    correlated. {(d)} Particle-wall collisions result in a momentum transfer in the counter spin
    wise direction \cite{TYRGL05}, the resulting edge flow propagation causes chirality reversal.  }
  \label{fig:correlations}
\end{figure*}


The $\mathbb{C+}\rightarrow\mathbb{C-}$ transition clearly has a continuous nature (no abrupt
transition from spin wise to counter spin wise vorticity is observed), because as we can see is
mediated by a distinct and continuously changing $\mathbb{C}\pm$ phase, with several (instead of
one) vortexes of both signs (Figure~\ref{fig:flow_vort}~(b), see also Movie 2
\cite{suppl}). Nevertheless, one of the signs predominates in the $\mathbb{C\pm}$, except at the
critical point. Yet, identification of the $\mathbb{C}\pm$ chirality is unambiguous as the count of
vortexes in the system is higher than one. The transitions are graphically described in
Figure~\ref{fig:phase_map} (b), which depicts the global vorticity
$\overline\omega=\overline{\omega}_++\overline{\omega}_-$ in the $(\phi, T_t)$ parameter space.



Interestingly, it is not trivial that there should be a chiral transition as the one described here,
nor that this transition should be necessarily continuous. For this reason, the results in
Figures~\ref{fig:flow_vort},~\ref{fig:phase_map} all lead us to consider that at least two different
mechanisms compete in the phase behavior, balancing each other to different extents in the
intermediate phase. Also, as we know, chirality emerges from an asymmetric part of the stress tensor
\cite {BSAV17} that is absent in regular fluids. One of the contributions to this asymmetric stress
tensor is proportional to the odd viscosity \cite{A98}. This odd viscosity, as other transport
coefficients, should emerge from the microscopic state structure of the system; i.e., the properties
of the non-equilibrium velocity distribution function \cite{CC70,VSK14}.

On the other hand, it is evident, from Figures~\ref{fig:distributions} (c), (d), that the
distribution functions show in general significant deviations from the Maxwellian, which are more
apparent for weak driving. These deviations off the Maxwellian have already been qualitatively
described detected in previous works on chiral fluids (see for instance \cite{WRDD18}). However, in
order to provide a more precise and quantitative description, one has to resort to the properties of
the cumulants of the distribution function, which additionally provide information on the
statistical correlations present in the system \cite{VSK14}. Surprisingly, the distribution function
cumulants and the statistical correlations they describe have so far remained unreported in the
context of chiral flows. Thus, in order to investigate on the origin of the mechanisms producing the
observed chiral transitions, we look here at the first four relevant spin/velocity moments
(cumulants are defined so that they are null for an equilibrium Maxwellian distribution function
\cite{VSK14}, see Appendix~\ref{AppendixB} details on their derivation),

\begin{align}
  & a_{20}^{(0)}(r) = \frac{1}{2}\left(\frac{1}{2}\frac{\langle V^4\rangle}{\langle
    V^2\rangle^2}-1\right),\nonumber\\
  & a_{02}^{(0)}(r) = \frac{1}{2}\left(\frac{1}{2}\frac{\langle W^4\rangle}{\langle W^2\rangle^2}-1\right) \nonumber\\
  & a_{11}^{(0)}(r) = \frac{1}{2}\left(\frac{\langle V^2W^2\rangle}{\langle
    V^2\rangle\langle W^2\rangle}-1\right),  \nonumber\\
  &  a_{00}^{(b)}(r)=
    \frac{3}{2}\frac{\langle(\mathbf{v}\times\mathbf{w})\cdot\mathbf{\hat
    e}_\varphi\rangle}{\sqrt{\langle V^2\rangle\langle W^2\rangle}}.
    \label{bend}
\end{align}

The cumulant $a_{00}^{(b)}$ in Equation~\ref{bend} (the unit vector $\mathbf{\hat e}_\phi$ denotes
the (counter clockwise) azimuthal direction) can be regarded as the analog of the bend coefficient
in the stress tensor of liquid crystals \cite{CL00}, and quantifies velocity-spin correlations in
the azimuthal direction (see Appendix~\ref{AppendixB} for more details). Notice that, according to
its definition in~\eqref{bend}, $a_{00}^{(b)}>0$ when particles tend to orbit performing a rotation
in the same sense as particles spin (i.e., the $\mathbb{C}_+$ phase), whereas for $a_{00}^{(b)}<0$
the fluid would orbit in opposite sense to particle spin (the $\mathbb{C}_-$ phase). Non chiral
fluids or with chirality changing locally would display vanishing $a_{00}^{(b)}$. In this way, we
take into account the inherent anisotropy of the system. Also, $a_{11}^{(0)}$ measures correlations
of the spin-velocities moduli. It is also relevant to consider the translational/rotational kinetic
energy fluctuations ratio, $T_t/T_r^*$, since this ratio indicates the degree of particle spin
synchronization; i.e., when particles rotate at nearly the same rate (spins \textit{synchronize}),
$T_r^*$ tends to be small against $T_t$ and thus $T_t/T_r^*$ increases.

We plot in Figure~\ref{fig:correlations} (a) the trends of the cumulants $a_{11}^{(0)}$,
$a_{00}^{(b)}$, together with $T_t/T_r^*$ (results for cumulants $a_{20}^{(0)}$ and $a_{02}^{(1)}$
do not display a clear trend except that in general they are not small, see SI Matherials and
Methods). Results for $a_{11}^{(0)}$ (blue symbols) and $a_{00}^{(b)}$ (inset) reveal indeed that
our system features correlations between rotations and translation. This coupling can yield the
$\mathbb{C}+$ chirality, as sketched in Figure~\ref{fig:correlations}~(c). In particular, it is
interesting to note that the sign of the bend cumulant $a_{00}^{(b)}$ evolves from negative for low
$\overline{T}_t$ to positive for high $\overline{T}_t$, thus determining the changes of sign of the
global vorticity $\overline\omega$. Also, the speed-spin correlations ($a_{11}^{(0)}$) mostly vanish
at higher $\overline{T}_t$, when $\mathbb{C}-$ chirality is predominant, with a fast decreasing
interval that roughly coincides with the transitional $\mathbb{C}\pm$ chirality. This correlation
decay is besides accompanied by a notable drop (of 2 orders of magnitude) of spin kinetic energy
fluctuations relative to translational energy fluctuations (set of red points, right Y axis in
Figure~(\ref{fig:correlations}~a), as the system increases its activity (higher $\overline{T}_t$) and
approaches the $\mathbb{C}-$ chiral state. This signals a process of spin synchronization that,
apparently, allows for the decay of the $\mathbb{C}+$ chiral mode. In summary, measurements indicate
that correlations between rotation and translations at the level of the particle dynamics have
distinctive features, according to the observed chiral phases.

The absence of correlations would not be the only factor involved in the build up of the
$\mathbb{C}-$ mode.
In fact, it is known that heat dissipation at the boundaries plays a significant role in flow
chirality reversal as well \cite{WRDD18}.  In this sense, we observe that, in the $\mathbb{C}\pm$
mode, counter spin wise vortexes tend to become first predominant in the boundaries (see
Figure~\ref{fig:flow_vort}~(b) and Movie 2 \cite{suppl}), which could indicate that in effect the
$\mathbb{C}_-$ mode builds up from the boundaries. This effect has been already detected but not
directly quantified yet. Thus, in Figure~\ref{fig:correlations}~(b), we plot
$\mathrm{d}T_t/\mathrm{d}r$ vs. $T_t$ as measured at the boundaries (main panel) and vs. $r$
(inset), since the normal heat flux is proportional to -$\mathrm{d}T_t/\mathrm{d}r$ \cite{CC70}. We
find that the heat flux undergoes a significant increase for high $\overline{T_t}$, coinciding with
the emergence of the $\mathbb{C}\pm$ and, afterwards, the $\mathbb{C}-$ mode. The increase of heat
dissipation at the boundaries indicates that wall-particle collisions (which are inelastic
\cite{VSK14}) are more frequent. Due to a mechanism of tangential friction (roughness) on impact,
this favors counter spin wise chirality after collision with the outer walls, see diagram in
Figure~\ref{fig:correlations}~(d), and thus the subsequent flow chirality reversal.

However, it is important to remark that heat dissipation is also proportional to particle collision
frequency \cite{VSK14}, which notably drops at low densities. Therefore, the chirality reversal at
low densities, as illustrated in Figure~\ref{fig:phase_map}~(b) (for instance in the transitions
found in the vertical lines at low densities) can only be explained by means of the spin
synchronization mechanism that according to our observations is also strong at low densities (see
low density series data in Figure~\ref{fig:correlations}). Moreover, we have found that the
inversion occurs also for boundary-frictionless systems, as reported in Movie 3 \cite{suppl}
material, which reinforces the idea that transitions build up of changes in the statistical
correlations and that boundaries are not necessarily involved in these processes (for more details
on this question, see theoretical analysis in Appendix A).

\section{Conclusion}
\label{conclusion}

In summary, we have described a system of active spinners with a rich and non-trivial dynamical
behavior, whose stationary flows feature three different chiral modes. The transition between these
modes is controlled by the interplay of spin-velocity correlations and boundary heat dissipation, as
we have shown; this interplay being controlled by kinetic energy fluctuations (primary role ) and
particle density (secondary role). In particular, it is interesting to notice that the main chiral
mode $\mathbb{C}+$ (for which fluid rotation is particle spin wise) is suppressed via a mechanism of
spin synchronization and bend coefficient reversal, eventually enhanced by heat dissipation at the
boundaries. 


Let us remark here that the fact that changes in the details of the coupling between particle
translations and rotations can induce changes in the average sign of particle chirality, which in
turn induces a transition in the sense of rotation of the chiral flow has been already been studied
in detail (see for instance \cite{GAS17,KS18,BLB20,ZYY20,ZPIBV16}. However, we report now a very different
situation. That is, by means of experimental evidence and theoretical analysis, we observed that a
change in the statistical correlations can also yield chiral flow transitions, these transitions
\textit{not} being mediated by a change in the sign of particle average chirality (particle spin
sign in the case of rotors), which actually remains here constant for all transitions.

Therefore, we report here a strong experimental evidence that the sign of chiral flow is not
necessarily determined by the type of particle chirality. In fact, it is statistical correlations
between particle velocity and spin that directly determine the sense of rotation of the chiral
flow. We think this result is also very relevant since, as a consequence, a variety of different chiral flows
emerge now as accessible for constant chirality particles as well.

Furthermore, the transitions between the observed chiral modes are defect-mediated and continuous
(due to the new and intermediate complex chiral phase $\mathbb{C}\pm$), as opposed the transitions
previously observed \cite{WRDD18}. To be more precise and according to our results, flow chirality
defects appear gradually (here, in the form of counter spinwise vortexes), thus yielding an
intermediate and previously not reported complex chiral phase, which we denoted as $\mathbb{C}\pm$
(see Figure~\ref{fig:flow_vort} a-c). This result is also relevant since it is in close analogy to
the scenario of transitions in other systems, such as semiflexible filament bundles \cite{SL21}, or
two-dimensional crystals \cite{S88}. Thus, clear links between chiral flow transitions and other
systems emerge now, in a broader context in soft matter.

Finally, our results illustrate that the set of transport coefficients inherent to chiral fluids,
including (but not only) odd viscosity \cite{A98,BSAV17}, would be controlled by the microscopic
behavior (i.e., the non-equilibrium distribution function features) of the system, just in the same
way as transport coefficients for regular fluids are \cite{CC70}. Future development of a kinetic
theory for this chiral of fluids would thus provide insight on their transport phenomena
properties. We expect to fulfill this in future work.

\appendix{

\section{Statistical mechanics of the two-dimensional chiral fluid}
\label{AppendixB}
 
\subsection{Properties of the kinetic equation}
\label{subsec:KE_prop}
Let us now consider the global features that the kinetic equation should have for our
system.

\begin{itemize}
\item First of all, particles display thermal-like translational movement (the concept of
  thermal-like movement for macroscopic particles was, probably, first instroduced by Kanatani, see
  \cite{K79}. Thus, the kinetic equation should be provided with an external forcing term
  $\mathbf{F}^\mathrm{th}$ that is acting stochastically each particle. This kind of forcing is
  usually denoted in non-equilibrium statistical mechanics as \textit{stochastic thermostat}
  \cite{EM90}. In this case we consider a stochastic themostat in the form of a white noise
  \cite{BB88,VK97}, which is defined by the relations
  $\langle\mathbf{F}^\mathrm{th}(t)\rangle=\mathbf{0},\:\langle F_i(t)
  F_j(t')\rangle=m^2\xi_0^2\delta_{ij}\delta(t-t')$, and $\xi_0^2$ is the forcing intensity. Notice
  that the forcing only correlates a particle with itself; i.e., there is no forcing between
  different particles (i.e., $i\neq j$).
\item Secondly, let us point out that particle activity emerges within the rotational degrees of
  freedom. This activity is inherently chiral since it is rooted in the geometric configuration of
  the particle, which here has \textit{chirality}. We will denote particles geometric chirality as
  $\mathbb{C}$, and rotational activity will be denoted as $\mathcal{A}(\mathbb{C}|\,\mathbf{w})$.
\item Lastly, particle collisions should also be chiral since momentum conservation laws imply, in
  general \cite{VSK14}, a back and forth transfer between the translational and the (here, chiral)
  rotational degrees of freedom. Additionally, it cannot be expected that active particle encounters
  will involve some energy exchange $\Delta E$ between the particles external dynamics and their
  internal degrees of motion. Therefore, collisions (encounters) between these active particles are
  \textit{inelastic}; i.e., they will not preserve energy.
\end{itemize}

Therefore, and if we only consider the case of a low density fluid, so that particle encounters occur by
pairs and their pre-collisional velocities are statistically uncorrelated; i.e., this implies a
Boltzmann-like equation \cite{CC70} with an inelastic \cite{BDKS98} and chiral collision operator

\begin{align}
  \label{eq:KE}
  \frac{\partial f}{\partial t}+\mathbf{v}\cdot\boldsymbol{\nabla}f -
  \frac{\xi_0^2}{2}\frac{\partial^2f}{\partial v^2} &+ \mathcal{A}(\mathbb{C}|\,\mathbf{w})[f]
                                                      \nonumber =
  \\ &J[\mathbf{r},\mathbf{v},(\mathbb{C}|\mathbf{w}); t|f,f].
\end{align}  We have tagged with the symbol
$\mathbb{C}$ each parity-violating term in Equation \eqref{eq:KE}. It is not the purpose of the
present work to analyze a specific collisional model that, in the case of active particles can be
rather involved \cite{CB06}. However, for our purpose, it will suffice here to consider for instance
the feature of hard collisions. In this case, the two-particle collision operator $J$ would be
of the form 

\begin{align}
  \label{coll_integral}
  &J[\mathbf{r},\mathbf{v}_1, (\mathbb{C}|\mathbf{w}_1); t|f,f] =
    \sigma\int\mathrm{d}\mathbf{v}_2\int\mathrm{d}\bm{\hat\sigma}
    \Theta(\mathbf{g}\bm{\cdot\hat\sigma})(\mathbf{g}\bm{\cdot\hat\sigma})
    \nonumber \\  & \times\left(b^{-1}[\Delta E, \mathbb{C}]-1\right)f(\mathbf{r}, \mathbf{v}_1,
                    \mathbf{w}_1,t) f(\mathbf{r}, \mathbf{v}_2, \mathbf{w}_2,t),
\end{align} where here $\mathbf{g}=\mathbf{v}_1-\mathbf{v}_2$, $\bm{\hat\sigma}$ is a unit
vector along the line joining centers of the colliding particles (with velocities and spins
$\mathbf{v}_i, \mathbf{w}_i$, respectively), and $\Theta$ is the Heaviside function. The function $b^{-1}$
operates on $f$ in such a way that yields the particles velocities \textit{before} the
collision. Thus, it can be denoted as \textit{restituting} operator \cite{BDKS98}. As we see, it
is both chiral and inelastic, and this has a fundamental effect on the dynamics of the system.

 Also, from Equation~\eqref{eq:KE} it is evident that there is a
mathematically complex coupling between the chirality in the spin activity and chirality in the
collision operator (which has a non-trivial integro-differential form), and as a result, it cannot
be asserted that chirality of the flow field $\mathbf{u}$ will yield a flow vorticity whose
chirality sign will necessarily coincide with that of $(\mathbb{C}|\mathbf{w})$ (vorticity is
defined here as $\omega(r)=(1/2)\epsilon_{ij}\partial_iu_j$, where $\epsilon_{ij}$ is the 2D
Levi-Civita symbol).

The question of the emergence of a fluid vorticity from a chiral flow pattern and the direction of
this vorticity is the fundamental question in the field of chiral active matter (see for instance
the last paragraph of section B2, page 16 in \cite{BFMR22}; or also references
\cite{TYRGL05,ZYSOS22} for instance, and most of the other ones in the bibliography) and the
analysis above vows to illustrate the fact that this question is not trivial and cannot depend on
the sign of particle chirality/activity or boundary conditions alone. Rather, it is a combination of
those together with the collisional properties of particle-particle collisions which will determine:
a) if chiral particles will produce a chiral flow at all; and b) if the flow chirality has the same
or opposite sign as particle chirality.

Furthermore, chirality of the system should emerge in the particle distribution itself, as we have
explained. Therefore, the distribution function should contain variables that are able to express
this parity violations taking place in the system. We analyze this in the next subsection.



\subsection{Particle velocities distribution function}
\label{subsec:fdistr_prop}

We will consider in this work stationary states only (distribution functions are time-independent),
thus, $\partial/\partial t=0$ in \eqref{eq:KE}. Additionally, if we consider small spatial
gradients, a normal solution for $f$ \cite{BDKS98} should be attainable, i.e., all of the space
dependence of the distribution function can be expressed through the relevant fields \cite{B73}. In
this case, this implies
$f(\mathbf{r},\mathbf{v},(\mathbb{C}|\mathbf{w}))=f(n,(\mathbb{C}|\mathbf{u}(\mathbf{r})),(\mathbb{C}|\mathbf{\Omega})(\mathbf{r}))$,
which shows that chirality will be transmitted from particle to average fields level, and thus the
chiral flow observed in Figure 1.

Moreover:
\begin{itemize}
\item The distribution function $f(r,\mathbf{v},\mathbf{w})$ will inherently have also
  chirality. Thus, a plausible way to describe them is through the a vectorial coupling of
  $\mathbf{v}$ and $\mathbf{w}$. For instance, the magnitude
  $\cos\alpha\equiv(\mathbf{v}\times\mathbf{w})\cdot\mathbf{\hat
    e}_\phi/|\mathbf{v}\times\mathbf{w}|$ would feature these properties.
\item $f(r,\mathbf{v},\mathbf{w})$ should be sensitive to the sign of $\cos\alpha$; since for
  $\cos\alpha<0$ the particle spin-velocity correlation is clockwise, thus favoring clockwise
  particle circulation, and thus clockwise flow vorticity. Conversely, for $\cos\alpha>0$, counter
  clockwise flow vorticity is favored. In this way, we guarantee that $f$ has the ability to display
  different flow vorticity behaviors for the same particle chirality $(\mathbb{C}|\mathbf{w})$.
\end{itemize}

We define the relevant average fields, particle density $n$, fluid flow $\mathbf{u}$ and average particle spin $\boldsymbol{\Omega}$

  \begin{equation}
    \begin{split}
      &n(r) = \int\mathrm{d}\mathbf{v}\int\mathrm{d}\mathbf{w}f(r,\mathbf{v},\mathbf{w}), \\
      &\mathbf{u}(r)=(1/n(r))\int\mathrm{d}\mathbf{v}\int\mathrm{d}\mathbf{w}\,f(r,\mathbf{v},\mathbf{w})\mathbf{v},
      \\
      &\boldsymbol\Omega(r)=(1/n(r))\int\mathrm{d}\mathbf{v}\int\mathrm{d}\mathbf{w}\,f(r,\mathbf{v},\mathbf{w})
      \mathbf{w},
    \end{split}
    \label{fields}
  \end{equation}  and $T_t$ (translational kinetic energy fluctuations) and $T_r$ (rotational kinetic energy)
  \begin{equation}
    \begin{split}
      &T_t(r) =\frac{m_p}{2n(r)} \int\mathrm{d}\mathbf{v}\int\mathrm{d}\mathbf{w}f(r,\mathbf{v},\mathbf{w}) (\mathbf{v}-\mathbf{u}(r))^2, \\
      &T_r(r) =\frac{I}{2n(r)}
      \int\mathrm{d}\mathbf{v}\int\mathrm{d}\mathbf{w}f(r,\mathbf{v},\mathbf{w}) w^2,
    \end{split}
    \label{temps}
  \end{equation}
  with $w=w_z$ (in our 2D system) and $m_p, I$ are particle mass and moment of inertia, respectively. We also define
  the spin kinetic energy fluctuations

 \begin{equation}
   T_r^*(r)=\frac{I}{2n(r)}
   \int\mathrm{d}\mathbf{v}\int\mathrm{d}\mathbf{w}f(r,\mathbf{v},\mathbf{w}) 
   (\mathbf{w}-\bm\Omega)^2.
 \end{equation} The scale of $T_t$ and $T_r^*$ in our system will in general be
 different.

 For convenience, we define $\mathbf{V}\equiv\mathbf{v}-\mathbf{u(r)}$,
 $\mathbf{W}\equiv\mathbf{w}-\bm{\Omega(r)}$; $V=(\mathbf{V}\cdot\mathbf{V})^{1/2}$,
 $W=(\mathbf{W}\cdot\mathbf{W})^{1/2}$. Let us also define the bivariate Maxellian

 \begin{equation}
   \begin{split}
     &f_{2M}(r,V^*,W^*) = \\
     &n(r)\left(\frac{m_p}{2\pi T_t(r)}\right) \left(\frac{I}{2\pi T_r^*(r)}\right)^{1/2}
     \mathrm{e}^{-\frac{V^2}{2T_t/m_p}}\mathrm{e}^{-\frac{W^2}{2T^*_r/I}} \\
     &\equiv n(r) \left(\frac{m_p}{2\pi T_t(r)}\right)\left(\frac{I}{2\pi
         T_r^*(r)}\right)^{1/2}\mathrm{e}^{-{V^*}^2}\mathrm{e}^{-{W^*}^2}.
     \label{f2M}
   \end{split}
 \end{equation}
 Henceforth, we use the notation:
 \begin{equation*}
   \begin{split}
     &{V^*}=V/(2T_t/m_p)^{1/2}\equiv V/\langle V\rangle^{1/2}, \\
     &{W^*}=W/(2T_r^*/I)^{1/2}\equiv W/\langle {W}^2\rangle^{1/2}
   \end{split}
 \end{equation*}
 for j-th powers of translational velocities and angular velocities, respectively, where we have
 taken into account \eqref{temps}.

 We can write the distribution function as a polynomial expansion around $f_{2M}(r,V^*,W^*)$.
 
 \begin{widetext}
   \begin{equation*}
     \begin{split}
       &f(r,V^*,W^*, \cos\alpha) = \\
       &f_{2M}(r,V^*,W^*) \sum_{j,k,\ell=0}^\infty
       a_{jk}^{(\ell)}\mathsf{L}_j^{(\ell+\ell_0)}({V^*}^2)\mathsf{L}_k^{(\ell+\ell_0')}({W^*}^2)\mathsf{P}_{\ell}(\cos\alpha)\,\mathsf{W}({V^*}^2,{V^*}^2,\cos\alpha) \\
       &= n(r) \frac{m}{2 T_t(r)}\left(\frac{I}{2 T_r^*(r)}\right)^{1/2}\sum_{j,k,\ell=0}^\infty
       a_{jk}^{(\ell)}\mathsf{L}_j^{(\ell+\ell_0)}({V^*}^2)\mathsf{L}_k^{(\ell+\ell_0')}({W^*}^2)\mathsf{P}_{\ell}(\cos\alpha)
       (\pi^{-3/2}\mathrm{e}^{-{V^*}^2}\mathrm{e}^{-{W^*}^2})\mathsf{W}({V^*},{W^*},\cos\alpha),
       \label{f_expansion}
     \end{split}
   \end{equation*} 
 \end{widetext}
 where $j,k,\ell$ are positive integers and $\ell_0, \ell_0'$ are (integer) constants. Symbols
 $\mathsf{L}_j^{(\ell)}(x)$, with $\ell$ being an integer, stand for the associated Laguerre
 polynomials of order $(j,\ell)$, and with $x\in[0,\infty]$. They are also commonly denoted, in the
 context of kinetic theory, as Sonine polynomials \cite{S80,NE98}. Also, $\mathsf{P}_\ell$ are the
 Legendre polynomials.
   
 The independent variable $\cos\alpha$ characterizes particle chirality, and is defined here as

 \begin{equation}
   \begin{split}
     \label{cos_alpha}
     &\cos\alpha\equiv \frac{\mathbf{v}\times\mathbf{w}}{|\mathbf{v}\times
       \mathbf{w}|}\cdot\mathbf{\hat e}_\varphi=w_z \frac{v_y\mathbf{\hat e}_x-v_x\mathbf{\hat
         e}_y}{|\mathbf{v}||\mathbf{w}|}\cdot\mathbf{\hat e}_\phi \\
     &=\frac{w_z}{w} \left(\frac{v_y\mathbf{\hat e}_x-v_x\mathbf{\hat
           e}_y}{v}\right)\cdot\left(\frac{-y\mathbf{\hat e}_x+x\mathbf{\hat e}_y}{r}\right) \\
     &=-\mathrm{sg}(w_z)\frac{\mathbf{v}\cdot\mathbf{r}}{r\,v}=-\mathrm{sg}(w)\frac{v_r}{v}.
   \end{split}
 \end{equation}

 Notice, from \eqref{cos_alpha}, that determination of $\cos\alpha$ involves also knowledge of
 particle position and thus, it will be considered henceforth as an integration variable that is
 independent of reduced particle velocity ($V^*$) and spin ($W^*$). Also, notice that these three
 features combined imply that in our system $\cos\alpha<0$ is to be expected for a flow whose
 vorticity is counter clockwise, and hence flow rotation has the same as particles spin. We will
 denote this situation as $\mathbb{C}_+$ (or spinwise, SW) chiral flow phase. Conversely, for
 $\cos\alpha>0$ yields clockwise vorticity; i.e., vorticity is counter spinwise (CSW), which we
 denote as $\mathbb{C}_-$.

 The product of associated Laguerre and Legendre polynomials both configure the set of orthogonal
 polynomials
 $\mathsf{L}_j^{(\ell)}({V^*}^2)\mathsf{L}_k^{(\ell)}({W^*}^2)({V^*W^*})^{2\ell}
 \mathsf{P}_{\ell}(\cos\alpha)$ in \eqref{f_expansion}. Use of the shifted Legendre polynimials
 \cite{nist,VSK14} is not justified here since, as we said, particle chirality depends on the sign
 of $\cos\alpha$. The weight function $\mathsf{W}({V^*},{W^*},\cos\alpha)$ is determined so that
   
 \begin{equation*}
   \begin{split}
     \int\mathrm{d}\mathbf{V}^*\int\mathrm{d}\mathbf{W}^*\int &\mathrm{d}(\cos\alpha)\mathsf{W}(V^  *,W^*,\cos\alpha) \\
     &(\pi^{-3/2}\mathrm{e}^{-{V^*}^2}\mathrm{e}^{-{W^*}^2}/W^*)
   \end{split}
 \end{equation*}
 \begin{equation*}
   \begin{split}
     =\int\mathrm{d}\mathbf{V}^*\int\mathrm{d}\mathbf{W}^*\int\mathrm{d}(\cos\alpha)\mathsf{w_L}^{(\ell)}({V^*}^2)\mathsf{w_L}^{(\ell)}({W^*}^2)\mathsf{w_P}(\cos\alpha)
   \end{split}
 \end{equation*}

 where $\mathsf{w_L}^{(\ell)}(x)\equiv e^{-x}x^{\ell},\, \mathsf{w_P}(x)\equiv 1$ are the weight
 functions of the associated Laguerre and Legendre polynomials respectively \cite{nist}, with
 respect to a variable $x$. Taking into account that
 $(m_p/2T_t(r))\int\mathrm{d}\mathbf{V}=\int\mathrm{d}\mathbf{V^*}=2\pi\int
 V^*\mathrm{d}V^*=\pi\int_0^\infty\mathrm{d}{V^*}^2$,
 $(I/2T_r^*(r))^{1/2}\int\mathrm{d}\mathbf{W}=2\int_
 0^\infty\mathrm{d}{W^*}=\int_0^\infty(1/W^*)\mathrm{d}{W^*}^2$, we obtain
 $\mathsf{W}(V^*,W^*,\cos\alpha)=\pi^{1/2}W^*(V^*W^*)^{2\ell}$.

 Therefore, for products of the type
 $\left[L_j^{(\ell+\ell_0)}({V^*}^2)L_k^{(\ell+\ell_0')}({W^*}^2)P_\ell(\cos\alpha)\right]\times
 f(r,V^*,W^*,\cos\alpha)$, we obtain
 \begin{widetext}
   \begin{equation*}
     \begin{split}
       & \int\mathrm{d}\mathbf{V}\int\mathrm{d}\mathbf{W}\int_{-1}^1\mathrm{d}(\cos\alpha)
       \left[L_j^{(\ell+\ell_0)}({V^*}^2)L_k^{(\ell+\ell_0')}({W^*}^2)P_\ell(\cos\alpha)\right]
       f(r,V^*,W^*, \cos\alpha) \\
       & = n(r)\sum_{j,k,\ell=0}^\infty\int_0
       ^\infty\mathrm{d}{V^*}^2\int_0^\infty\mathrm{d}{W^*}^2\int_
       {-1}^{+1}\mathrm{d}(\cos\alpha)\left(\mathsf{L}_j^{(\ell)}({V^*}^2)\mathsf{L}_k^{(\ell)}({W^*}^2)\mathsf{P}_{\ell}(\cos\alpha)
       \right)\left(\mathsf{L}_{j'}^{(\ell')}({V^*}^2)\mathsf{L}_{k'}^{(\ell')}({W^*}^2)\mathsf{P}_{\ell'}(\cos\alpha)\right) \\
       &\times\mathrm{e}^{{-V^*}^2-{W^*}^2}({V^*}{W^*})^{2\ell} \\
       & =n(r)\sum_{j,k,\ell=0}^\infty a_{jk}^{(\ell)}\frac{\Gamma(j+\ell+1)}{j!}
       \frac{\Gamma(k+\ell+1)}{k!}  \frac{2}{2\ell+1}\delta_{j,j'}\delta_{k,k'}\delta_{\ell,\ell'} =
       n(r) a_{jk}^{(\ell)}\frac{\Gamma(j+\ell+\ell_0+1)}{j!}  \frac{\Gamma(k+\ell+\ell_0'+1)}{k!}
       \frac{2}{2\ell+1},
       \label{norm_ct}
     \end{split}
   \end{equation*} 
 \end{widetext}
 where we have taken into account the orthogonality conditions for the associated Laguerre and
 Legendre polynomials \cite{nist}.

 In order to determine the constants $\ell_0, \ell_0'$ we determine the integral in~\eqref{norm_ct},
 for $j=1, k=0,\ell=0$, taking into account also~\eqref{temps}

 \begin{widetext}
   \begin{equation*}
     \begin{split}
       \int\mathrm{d}\mathbf{V}\int\mathrm{d}\mathbf{W}&\int\mathrm{d}(\cos\alpha)L_1^{(\ell_0)}({V^*}^2)L_0^{(\ell_0')}({W^*}^2)P_0(\cos\alpha)
       f(r,\mathbf{v},\mathbf{w})=\int\mathrm{d}\mathbf{V}\int\mathrm{d}\mathbf{W}\int\mathrm{d}(\cos\alpha)(1+\ell_0-{V^*}^2)
       f(r,\mathbf{v},\mathbf{w}) \\
       &= \left(1+\ell_0-\frac{\langle V^2\rangle}{2T_t(r)/m_p}\right)n(r),
       \label{moment1a}
     \end{split}
   \end{equation*} 
 \end{widetext}
 (where we have used
 $\langle\dots\rangle=
 (1/n(r))\int\mathrm{d}\textbf{V}\int\mathrm{d}\textbf{W}\int\mathrm{d}(\cos\alpha)(\dots)
 f(r,\mathbf{v},\mathbf{w})$) which, by comparing with \eqref{norm_ct} leads to

\begin{equation}
  2\Gamma(\ell_0+2)\Gamma(\ell_0'+1)a_{10}^{(0)} =
  1+\ell_0-\frac{\langle V^2\rangle}{2T_t(r)/m_p}.
  \label{a10_prev}
\end{equation} 

The equation above makes evident the convenient choice $\ell_0=0$, for which \eqref{a10_prev} yields

\begin{equation}
  2\Gamma(\ell_0'+1)a_{10}^{(0)} = 1-\frac{\langle V^2\rangle}{2T_t/m_p}=0, \quad\Rightarrow\quad a_{10}^{(0)}=0,
  \label{a10}
\end{equation} since the definition of average translational kinetic energy $T_t$
in~\eqref{temps} implies that $(m/2)\langle V^2\rangle=T_t$. We proceed analogously to obtain $a_{01}^{(0)}=0$ for $\ell_0'=0$.


Moreover, the choice $\ell_0, \ell_0'=0$ implies that all $a_{j,k}^{(\ell)}=0$ except for the zeroth
order contribution $a_{00}^{(0)}$, if the particle distribution function is a 2D Maxwellian
(i.e.,$f=f_{2M}$). Therefore, these constants (usually called cumulants or Sonine coefficients in
the context of kinetic theory \cite{VSK14}) measure deviations from the Maxwellian (i.e., from the
thermodynamic equilibrium state).

Thus, the expansion series for our distribution function would write as

\begin{equation}
  \begin{split}
    &f(r,\mathbf{V},\mathbf{W}) = \\
    &f_{2M}(r,V,W) \sum_{j,k,\ell=0}^\infty
    a_{jk}^{(\ell)}L_j^{(\ell)}({V^*}^2)L_k^{(\ell)}({W^*}^2)(\pi^{1/2}W^*) \\
    &({V^*}^2{W^*}^2)^{\ell}P_{\ell}(\cos\alpha).
  \end{split}
  \label{fdistr_f}
\end{equation}

Notice that for $\ell_0,\ell_0'=0$ the first three associated Laguerre polynomials are
$L_0^{\ell}(x)=1$, $L_1^{(\ell)}(x)=(1+\ell)-x$, $L_2^{(\ell)}(x)=x^2/2-(2+\ell)x+1$. On the other
hand, the first two Legendre polynomials are $P_0(y)=1$, $P_1(y)=y$. Thus, by replacing
$x\to {V^*}^2,{W^*}^2$ and $y\to \cos\alpha$ and repeating the procedure
in~\eqref{moment1a},~\eqref{a10_prev} for the polynomial combinations
$L_2^{(0)}({V^*}^2)L_0^{(0)}({W^*}^2)P_0(\cos\alpha)$,
$L_0^{(0)}({V^*}^2)L_2^{(0)}({W^*}^2)P_0(\cos\alpha)$,
$L_1^{(0)}({V^*}^2)L_1^{(0)}({W^*}^2)P_0(\cos\alpha)$ and
$L_0^{(1)}({V^*}^2)L_0^{(1)}({W^*}^2)P_1(\cos\alpha)$ we obtain, respectively, the first four
cumulants

\begin{equation}
  \begin{split}
    &a_{20}^{(0)}(r) = \frac{1}{2}\left(\frac{1}{2}\frac{\langle V^4\rangle}{\langle
        V^2\rangle^2}-1\right),\hspace{0.35em}   \\
    &a_{02}^{(0)}(r) = \frac{1}{2}\left(\frac{1}{2}\frac{\langle
        W^4\rangle}{\langle W^2\rangle^2}-1\right), \hspace{0.35em} \\
    &a_{11}^{(0)}(r) = \frac{1}{2}\left(\frac{\langle V^2W^2\rangle}{\langle
        V^2\rangle\langle W^2\rangle}-1\right), \hspace{0.35em} \\
    &a_{00}^{(1)}(r)=\frac{3}{2}\left\langle\frac{\mathbf{v}\times\mathbf{w}\cdot\mathbf{\hat
          e}_\phi}{|\mathbf{v}\times\mathbf{w}|}\right\rangle,
  \end{split}
\end{equation} where for the expression of $a_{00}^{(1)}$ we have taken into account our definition $\cos\alpha\equiv \langle(\mathbf{v}\times\mathbf{w})\cdot\mathbf{\hat
  e}_\phi/|\mathbf{v}\times\mathbf{w}|\rangle$.

The cumulant $a_{00}^{(1)}$ characterizes particle translation-spin correlations. However, notice
that for characterization of particle translation-spin correlations we use instead the slightly
different coefficient $a_{00}^{(b)}$

\begin{equation}
  \label{a_00b}
  a_{00}^{(b)} = \frac{3}{2}{\left\langle(\mathbf{v}^*\times\mathbf{w}^*)\cdot\mathbf{\hat
        e}_\varphi\right\rangle},
\end{equation} (with $\mathbf{v}^*\equiv \mathbf{v}/\sqrt{\langle V^2\rangle}$
and $\mathbf{w}^*\equiv \mathbf{w}/\sqrt{\langle W^2\rangle}$) which retains analogous
statistical information
on the relative orientation of particle translations and spin, while holding a closer
analogy to the bend coefficient used in elasticity theory \cite{CL00}. For this reason, we used $a_{00}^{(b)}$ to characterize translation-spin correlations in Figure 4 (a) in the main file.

  \section{Materials and experimental methods}
  \label{AppendixA}
  
  We describe in this section the experimental methods. Additional experimental results and data,
  together with a description of experiments movies are provided in the Supplementary Material file
  and in an external data set \cite{suppl}.

  The particles we used in our experiments are 3D-printed polylactic acid (PLA) disks of diameter
  $\sigma=72.5~\mathrm{mm}$ and height $h=6~\mathrm{mm}$ and they have 14 oblique blades that
  generate a clock-wise spinning as an air flow passes through the disk (see
  Figure~\ref{fig:distributions}~(a)). The rotors are located on top of a perforated steel
  sheet ($3~\mathrm{mm}$ diameter holes in a hexagonal pattern, with a $3~\mathrm{mm}$ spacing),
  this grid has been carefully leveled and is mounted on a box that guides an adjustable air
  current, generated by a fan and homogeneized by means of a polyurethane foam layer; uniformity of
  the airflow is verified with an anemometer and local deviations are found to be within $\pm 5~\%$
  of the averaged value; mean air flows ranged from $2.2$ to $3.2~\mathrm{m/s}$. Our disks are
  contained inside a circular PLA border of $725~\mathrm{mm}$ diameter.

  We have recorded a total of 120 experiments, carefully varying the parameters (density and air
  current), each realization consists of two movies (filmed with a Phantom VEO 410L high-speed
  camera). The first has a duration of $27.75~\mathrm{s}$ and is recorded at 900 frames per second,
  which allows for measuring particles spin. The second take has a duration of $\sim 100~\mathrm{s}$
  at 250 fps, and is used to calculate vorticity and the other relevant fields. Positions are
  calculated using a modified version of Crocker and Grier algorithm \cite{Crocker1996}, obtaining a
  spatial resolution of around $0.05~\%$ of the particle diameter. Meanwhile, spinner angular
  velocities were found using a custom method based on tracking blade luminosity, the accuracy in
  the measurement of frame-to-frame angular displacements is $\pm 3.5\times10^{-3}~\mathrm{rad}$.

} 

\begin{acknowledgments}
  We acknowledge funding from the Government of Spain through project Nos. FIS2016-76359-P,
  PID2020-116567GB-C22 and from the regional Extremadura Government through projects No. GR18079,
  GR21091 \& IB16087, IB20079 partially funded by the ERDF. A.R.-R. also acknowledges financial
  support from Conserjer\'ia de Transformaci\'on Económica, Industria, Conocimiento y Universidades
  de la Junta de Andaluc\'ia through post-doctoral grant no. DC 00316 (PAIDI 2020), co-funded by the
  EU Fondo Social Europeo (FSE).
\end{acknowledgments}
\bibliography{chiral_t}
\clearpage
\input{SM_input.tex}

\end{document}

%% file: SM_input.tex
\newpage

\section*{Supplementary figures and  experiments movies}

In this work we have recorded a total of 120 movies, covering a wide range of densities, going from
$\phi=0.03$ to $\phi=0.70$.  We could not represent this comprehensive experimental database. We
include here, however two additional figures, 3 supplementary movies and experimental data.

\begin{figure*}
  \centering \includegraphics[width=0.75\textwidth]{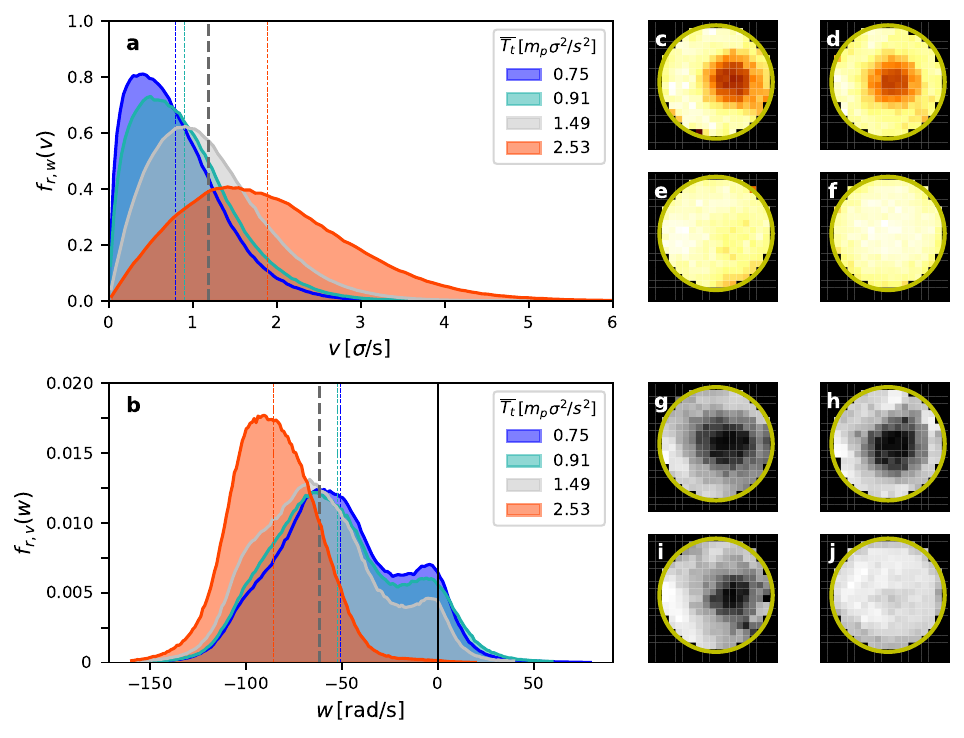}
  \caption{(a) Translational particle velocity distributions ($v$), $f_{r,w}(v)$ for a system with
    $\phi=0.55$ and four different values of $\overline{T_t}$. (b) particle spin distribution ($w$),
    $f_{r,v}(w)$. In (a)-(b), dashed vertical lines indicate the mean of each distribution function
    (follow color correspondence). In (b), the continuous vertical line indicates separation between
    $\omega<0$ and $\omega>0$. (c-f) Reduced particle speed fluctuations
    $U^*(x,y)\equiv v_\mathrm{th}(x,y)/V_0$. (g-j) Reduced average spin
    $\Omega^*(x,y)\equiv\left< w(x,y)\right>/w_0$.  In both $U^*(x,y)$ and $\Omega^*(x,y)$ color
    maps, brighter is higher and darker is lower; and black stands for $U^*(x,y),\Omega^*(x,y)=0$
    and white for $U^*(x,y), \Omega^*(x,y)=1$. $T_t$ is in units of $m_p\sigma^2/\mathrm{s}^2$. The
    system boundary is marked with a thick yellow line.}
  \label{fig:suppl_fig1}
\end{figure*}

\begin{figure*}
  \centering \includegraphics[width=0.4\textwidth]{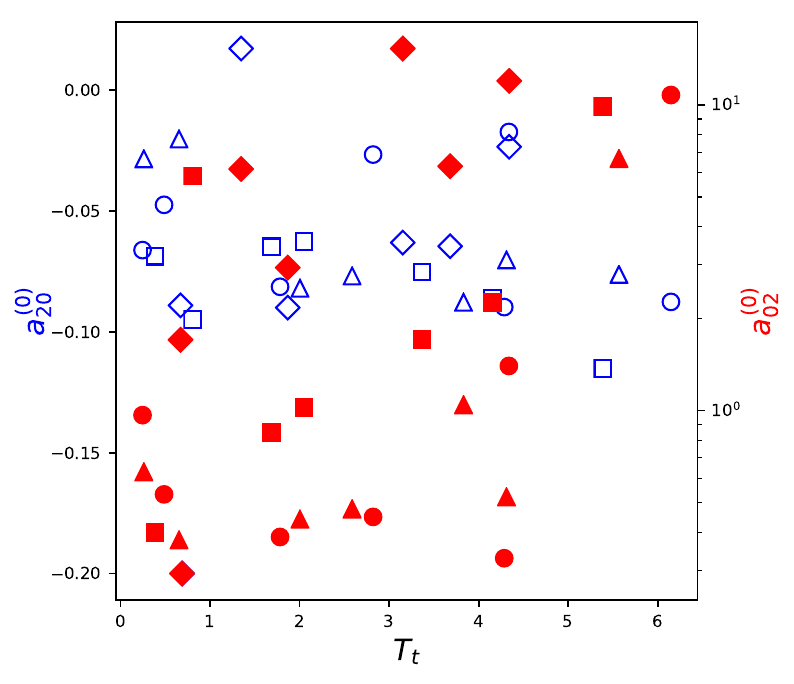}
    \caption{In this figure we plot, vs. $\overline{T_t}$, the cumulants $a_{20}^{(0)}$ (blue, $Y$
      axis at the left), $a_{02}^{(0)}$ (red, $Y$ axis at the right in logarithmic scale). Each
      point represents the value for a single experiment. We show that the systems are clearly not
      Maxwellian (values far from zero). Also, there is no clear trend of either $a_{20}^{(0)}$ nor      $a_{02}^{(0)}$ as a function of $\overline{T_t}$.}
    \label{fig:suppl_fig2}
  \end{figure*}

\begin{itemize}
\item \textbf{Additional Figures:} The first supplementary figure is analogous to Figure 1 in the
  main text, except that here the packing fraction is $\phi=0.55$. Figure~\ref{fig:suppl_fig1} in
  this Supplementary Material one can see that the bimodal behavior of the spin distribution
  function is stronger than in the lower density case (shown in the main text). Also, from panels
  \textbf{c-j} we now see that particle cover the entire area of the system.  The second
  supplementary figure (Figure~\ref{fig:suppl_fig2}) includes data for the cumulants defined in
  Eq. 2 (main text) that are not represented by Figure 4 in main text.
\item \textbf{Supplementary Movies 1 \& 2:} We included two additional movies in order to help the
  reader understand the most relevant results. In Movie 1, we show three configurations with
  constant density but an increasing thermalization level
  ($T_t=\{ 0.63, ~1.16, ~2.65 \} ~m_p (\sigma/\mathrm{s})^{2}$), this movie illustrate the chirality
  reversal caused by interaction with the system boundaries, we display the trajectories of three
  representative particles for each case. Movie 2 displays the evolution of the vorticity field for
  a sample experiment, there we show a situation of complex chirality with several vortexes of
  opposite directions, these vortexes evolve with time.
\item \textbf{Supplementary Movie 3:} Experiment movie for $N=3$. It can be seen by eye that the
  chiral flow inversion occurs without intervention of the boundaries. Hence, chiral flow inversion
  is not in general boundary-driven.
\end{itemize}

   \section*{Data availability}

   The data that support the plots within this paper and other findings of this study
   are available in the following web link:  \texttt{https://zenodo.org/record/4756796} .

   \section*{Codes availability}

   The particle tracking codes that were used for this work are available in the
   following web link: \newline\texttt{https://github.com/fvegar/blades}.


%% file: chiral_t.bbl
\begin{thebibliography}{47}%
\makeatletter
\providecommand \@ifxundefined [1]{%
 \@ifx{#1\undefined}
}%
\providecommand \@ifnum [1]{%
 \ifnum #1\expandafter \@firstoftwo
 \else \expandafter \@secondoftwo
 \fi
}%
\providecommand \@ifx [1]{%
 \ifx #1\expandafter \@firstoftwo
 \else \expandafter \@secondoftwo
 \fi
}%
\providecommand \natexlab [1]{#1}%
\providecommand \enquote  [1]{``#1''}%
\providecommand \bibnamefont  [1]{#1}%
\providecommand \bibfnamefont [1]{#1}%
\providecommand \citenamefont [1]{#1}%
\providecommand \href@noop [0]{\@secondoftwo}%
\providecommand \href [0]{\begingroup \@sanitize@url \@href}%
\providecommand \@href[1]{\@@startlink{#1}\@@href}%
\providecommand \@@href[1]{\endgroup#1\@@endlink}%
\providecommand \@sanitize@url [0]{\catcode `\\12\catcode `\$12\catcode
  `\&12\catcode `\#12\catcode `\^12\catcode `\_12\catcode `\%12\relax}%
\providecommand \@@startlink[1]{}%
\providecommand \@@endlink[0]{}%
\providecommand \url  [0]{\begingroup\@sanitize@url \@url }%
\providecommand \@url [1]{\endgroup\@href {#1}{\urlprefix }}%
\providecommand \urlprefix  [0]{URL }%
\providecommand \Eprint [0]{\href }%
\providecommand \doibase [0]{https://doi.org/}%
\providecommand \selectlanguage [0]{\@gobble}%
\providecommand \bibinfo  [0]{\@secondoftwo}%
\providecommand \bibfield  [0]{\@secondoftwo}%
\providecommand \translation [1]{[#1]}%
\providecommand \BibitemOpen [0]{}%
\providecommand \bibitemStop [0]{}%
\providecommand \bibitemNoStop [0]{.\EOS\space}%
\providecommand \EOS [0]{\spacefactor3000\relax}%
\providecommand \BibitemShut  [1]{\csname bibitem#1\endcsname}%
\let\auto@bib@innerbib\@empty
\bibitem [{\citenamefont {Wu}\ \emph {et~al.}(2011)\citenamefont {Wu},
  \citenamefont {Hosu},\ and\ \citenamefont {Berga}}]{WHB11}%
  \BibitemOpen
  \bibfield  {author} {\bibinfo {author} {\bibfnamefont {Y.}~\bibnamefont
  {Wu}}, \bibinfo {author} {\bibfnamefont {B.~G.}\ \bibnamefont {Hosu}},\ and\
  \bibinfo {author} {\bibfnamefont {H.~C.}\ \bibnamefont {Berga}},\ }\bibfield
  {title} {\bibinfo {title} {Microbubbles reveal chiral fluid flows in
  bacterial swarms},\ }\href@noop {} {\bibfield  {journal} {\bibinfo  {journal}
  {Proc. Natl. Acad. Sci. USA}\ }\textbf {\bibinfo {volume} {108}},\ \bibinfo
  {pages} {4147} (\bibinfo {year} {2011})}\BibitemShut {NoStop}%
\bibitem [{\citenamefont {Soni}\ \emph {et~al.}(2019)\citenamefont {Soni},
  \citenamefont {Bililign}, \citenamefont {Magkiriadou}, \citenamefont
  {Sacanna}, \citenamefont {Bartolo}, \citenamefont {Shelley},\ and\
  \citenamefont {Irvine}}]{SBSI19}%
  \BibitemOpen
  \bibfield  {author} {\bibinfo {author} {\bibfnamefont {V.}~\bibnamefont
  {Soni}}, \bibinfo {author} {\bibfnamefont {E.~S.}\ \bibnamefont {Bililign}},
  \bibinfo {author} {\bibfnamefont {S.}~\bibnamefont {Magkiriadou}}, \bibinfo
  {author} {\bibfnamefont {S.}~\bibnamefont {Sacanna}}, \bibinfo {author}
  {\bibfnamefont {D.}~\bibnamefont {Bartolo}}, \bibinfo {author} {\bibfnamefont
  {M.~J.}\ \bibnamefont {Shelley}},\ and\ \bibinfo {author} {\bibfnamefont
  {W.~T.~M.}\ \bibnamefont {Irvine}},\ }\bibfield  {title} {\bibinfo {title}
  {The odd free surface flows of a colloidal chiral fluid},\ }\href@noop {}
  {\bibfield  {journal} {\bibinfo  {journal} {Nat. Phys.}\ }\textbf {\bibinfo
  {volume} {15}},\ \bibinfo {pages} {1188} (\bibinfo {year}
  {2019})}\BibitemShut {NoStop}%
\bibitem [{\citenamefont {Tsai}\ \emph {et~al.}(2005)\citenamefont {Tsai},
  \citenamefont {Ye}, \citenamefont {Rodriguez}, \citenamefont {Gollub},\ and\
  \citenamefont {Lubensky}}]{TYRGL05}%
  \BibitemOpen
  \bibfield  {author} {\bibinfo {author} {\bibfnamefont {J.~C.}\ \bibnamefont
  {Tsai}}, \bibinfo {author} {\bibfnamefont {F.}~\bibnamefont {Ye}}, \bibinfo
  {author} {\bibfnamefont {J.}~\bibnamefont {Rodriguez}}, \bibinfo {author}
  {\bibfnamefont {J.~P.}\ \bibnamefont {Gollub}},\ and\ \bibinfo {author}
  {\bibfnamefont {T.~C.}\ \bibnamefont {Lubensky}},\ }\bibfield  {title}
  {\bibinfo {title} {A chiral granular gas},\ }\href@noop {} {\bibfield
  {journal} {\bibinfo  {journal} {Phys. Rev. Lett.}\ }\textbf {\bibinfo
  {volume} {94}},\ \bibinfo {pages} {214301} (\bibinfo {year}
  {2005})}\BibitemShut {NoStop}%
\bibitem [{\citenamefont {Banerjee}\ \emph {et~al.}(2017)\citenamefont
  {Banerjee}, \citenamefont {Souslov}, \citenamefont {Abanov},\ and\
  \citenamefont {Vitelli}}]{BSAV17}%
  \BibitemOpen
  \bibfield  {author} {\bibinfo {author} {\bibfnamefont {D.}~\bibnamefont
  {Banerjee}}, \bibinfo {author} {\bibfnamefont {A.}~\bibnamefont {Souslov}},
  \bibinfo {author} {\bibfnamefont {A.~G.}\ \bibnamefont {Abanov}},\ and\
  \bibinfo {author} {\bibfnamefont {V.}~\bibnamefont {Vitelli}},\ }\bibfield
  {title} {\bibinfo {title} {Odd viscosity in chiral active fluids},\
  }\href@noop {} {\bibfield  {journal} {\bibinfo  {journal} {Nat. Commun.}\
  }\textbf {\bibinfo {volume} {8}},\ \bibinfo {pages} {1573} (\bibinfo {year}
  {2017})}\BibitemShut {NoStop}%
\bibitem [{\citenamefont {Lei}\ and\ \citenamefont {Ni}(2019)}]{LN19}%
  \BibitemOpen
  \bibfield  {author} {\bibinfo {author} {\bibfnamefont {Q.-L.}\ \bibnamefont
  {Lei}}\ and\ \bibinfo {author} {\bibfnamefont {R.}~\bibnamefont {Ni}},\
  }\bibfield  {title} {\bibinfo {title} {Hydrodynamics of random-organizing
  hyperuniform fluids},\ }\href@noop {} {\bibfield  {journal} {\bibinfo
  {journal} {Proc. Natl. Acad. Sci. USA}\ }\textbf {\bibinfo {volume} {116}},\
  \bibinfo {pages} {22983} (\bibinfo {year} {2019})}\BibitemShut {NoStop}%
\bibitem [{\citenamefont {Farhadi}\ \emph {et~al.}(2018)\citenamefont
  {Farhadi}, \citenamefont {Machaca}, \citenamefont {Aird}, \citenamefont
  {Maldonado}, \citenamefont {Davis}, \citenamefont {Arratia},\ and\
  \citenamefont {Durian}}]{DAD18}%
  \BibitemOpen
  \bibfield  {author} {\bibinfo {author} {\bibfnamefont {S.}~\bibnamefont
  {Farhadi}}, \bibinfo {author} {\bibfnamefont {S.}~\bibnamefont {Machaca}},
  \bibinfo {author} {\bibfnamefont {J.}~\bibnamefont {Aird}}, \bibinfo {author}
  {\bibfnamefont {B.~T.}\ \bibnamefont {Maldonado}}, \bibinfo {author}
  {\bibfnamefont {S.}~\bibnamefont {Davis}}, \bibinfo {author} {\bibfnamefont
  {P.}~\bibnamefont {Arratia}},\ and\ \bibinfo {author} {\bibfnamefont
  {D.}~\bibnamefont {Durian}},\ }\bibfield  {title} {\bibinfo {title} {Dynamics
  and thermodynamics of air-driven active spinners},\ }\href@noop {} {\bibfield
   {journal} {\bibinfo  {journal} {Soft Matter}\ }\textbf {\bibinfo {volume}
  {14}},\ \bibinfo {pages} {5588} (\bibinfo {year} {2018})}\BibitemShut
  {NoStop}%
\bibitem [{\citenamefont {Nguyen}\ \emph {et~al.}(2014)\citenamefont {Nguyen},
  \citenamefont {Klotsa}, \citenamefont {Engel},\ and\ \citenamefont
  {Glotzer}}]{NKEG14}%
  \BibitemOpen
  \bibfield  {author} {\bibinfo {author} {\bibfnamefont {N.~H.~P.}\
  \bibnamefont {Nguyen}}, \bibinfo {author} {\bibfnamefont {D.}~\bibnamefont
  {Klotsa}}, \bibinfo {author} {\bibfnamefont {M.}~\bibnamefont {Engel}},\ and\
  \bibinfo {author} {\bibfnamefont {S.~C.}\ \bibnamefont {Glotzer}},\
  }\bibfield  {title} {\bibinfo {title} {Emergent collective phenomena in a
  mixture of hard shapes through active rotation},\ }\href@noop {} {\bibfield
  {journal} {\bibinfo  {journal} {Phys. Rev. Lett.}\ }\textbf {\bibinfo
  {volume} {112}},\ \bibinfo {pages} {075701} (\bibinfo {year}
  {2014})}\BibitemShut {NoStop}%
\bibitem [{\citenamefont {Zhang}\ \emph {et~al.}(2020)\citenamefont {Zhang},
  \citenamefont {Sokolov},\ and\ \citenamefont {Snezhko}}]{ZSS20}%
  \BibitemOpen
  \bibfield  {author} {\bibinfo {author} {\bibfnamefont {B.}~\bibnamefont
  {Zhang}}, \bibinfo {author} {\bibfnamefont {A.}~\bibnamefont {Sokolov}},\
  and\ \bibinfo {author} {\bibfnamefont {A.}~\bibnamefont {Snezhko}},\
  }\bibfield  {title} {\bibinfo {title} {Reconfigurable emergent patterns in
  active chiral fluids},\ }\href@noop {} {\bibfield  {journal} {\bibinfo
  {journal} {Nat. Commun.}\ }\textbf {\bibinfo {volume} {11}},\ \bibinfo
  {pages} {4401} (\bibinfo {year} {2020})}\BibitemShut {NoStop}%
\bibitem [{\citenamefont {Souslov}\ \emph {et~al.}(2017)\citenamefont
  {Souslov}, \citenamefont {{van Zuiden}}, \citenamefont {Bartolo},\ and\
  \citenamefont {Vitelli}}]{SZBV17}%
  \BibitemOpen
  \bibfield  {author} {\bibinfo {author} {\bibfnamefont {A.}~\bibnamefont
  {Souslov}}, \bibinfo {author} {\bibfnamefont {B.~C.}\ \bibnamefont {{van
  Zuiden}}}, \bibinfo {author} {\bibfnamefont {D.}~\bibnamefont {Bartolo}},\
  and\ \bibinfo {author} {\bibfnamefont {V.}~\bibnamefont {Vitelli}},\
  }\bibfield  {title} {\bibinfo {title} {Topological sound in active-liquid
  metamaterials},\ }\href@noop {} {\bibfield  {journal} {\bibinfo  {journal}
  {Nat. Phys.}\ }\textbf {\bibinfo {volume} {13}},\ \bibinfo {pages} {1091}
  (\bibinfo {year} {2017})}\BibitemShut {NoStop}%
\bibitem [{\citenamefont {Yang}\ \emph {et~al.}(2021)\citenamefont {Yang},
  \citenamefont {Zhu}, \citenamefont {Liu}, \citenamefont {Liu}, \citenamefont
  {Shi}, \citenamefont {Chen}, \citenamefont {Zheng}, \citenamefont {Ye},\ and\
  \citenamefont {Yang}}]{Yetal21}%
  \BibitemOpen
  \bibfield  {author} {\bibinfo {author} {\bibfnamefont {Q.}~\bibnamefont
  {Yang}}, \bibinfo {author} {\bibfnamefont {H.}~\bibnamefont {Zhu}}, \bibinfo
  {author} {\bibfnamefont {P.}~\bibnamefont {Liu}}, \bibinfo {author}
  {\bibfnamefont {R.}~\bibnamefont {Liu}}, \bibinfo {author} {\bibfnamefont
  {Q.}~\bibnamefont {Shi}}, \bibinfo {author} {\bibfnamefont {K.}~\bibnamefont
  {Chen}}, \bibinfo {author} {\bibfnamefont {N.}~\bibnamefont {Zheng}},
  \bibinfo {author} {\bibfnamefont {F.}~\bibnamefont {Ye}},\ and\ \bibinfo
  {author} {\bibfnamefont {M.}~\bibnamefont {Yang}},\ }\bibfield  {title}
  {\bibinfo {title} {Topologically protected transport of cargo in a chiral
  active fluid aided by odd-viscosity-enhanced depletion interactions},\
  }\href@noop {} {\bibfield  {journal} {\bibinfo  {journal} {Phys. Rev. Lett.}\
  }\textbf {\bibinfo {volume} {126}},\ \bibinfo {pages} {198001} (\bibinfo
  {year} {2021})}\BibitemShut {NoStop}%
\bibitem [{\citenamefont {Avron}(1998)}]{A98}%
  \BibitemOpen
  \bibfield  {author} {\bibinfo {author} {\bibfnamefont {J.~E.}\ \bibnamefont
  {Avron}},\ }\bibfield  {title} {\bibinfo {title} {Odd viscosity},\
  }\href@noop {} {\bibfield  {journal} {\bibinfo  {journal} {J. Stat. Phys.}\
  }\textbf {\bibinfo {volume} {92}},\ \bibinfo {pages} {543} (\bibinfo {year}
  {1998})}\BibitemShut {NoStop}%
\bibitem [{\citenamefont {Han}\ \emph {et~al.}(2020)\citenamefont {Han},
  \citenamefont {Kokot}, \citenamefont {Tovkach}, \citenamefont {Glatz},
  \citenamefont {Aranson},\ and\ \citenamefont {Snezhko}}]{HKTGAS20}%
  \BibitemOpen
  \bibfield  {author} {\bibinfo {author} {\bibfnamefont {K.}~\bibnamefont
  {Han}}, \bibinfo {author} {\bibfnamefont {G.}~\bibnamefont {Kokot}}, \bibinfo
  {author} {\bibfnamefont {O.}~\bibnamefont {Tovkach}}, \bibinfo {author}
  {\bibfnamefont {A.}~\bibnamefont {Glatz}}, \bibinfo {author} {\bibfnamefont
  {I.}~\bibnamefont {Aranson}},\ and\ \bibinfo {author} {\bibfnamefont
  {A.}~\bibnamefont {Snezhko}},\ }\bibfield  {title} {\bibinfo {title}
  {Emergence of self-organized multivortex states in flocks of active
  rollers},\ }\href@noop {} {\bibfield  {journal} {\bibinfo  {journal} {Proc.
  Natl. Acad. Sci. USA}\ }\textbf {\bibinfo {volume} {117}},\ \bibinfo {pages}
  {9706} (\bibinfo {year} {2020})}\BibitemShut {NoStop}%
\bibitem [{\citenamefont {van Zuiden}\ \emph {et~al.}(2016)\citenamefont {van
  Zuiden}, \citenamefont {Paulose}, \citenamefont {Irvine}, \citenamefont
  {Bartolo},\ and\ \citenamefont {Vitelli}}]{ZPIBV16}%
  \BibitemOpen
  \bibfield  {author} {\bibinfo {author} {\bibfnamefont {B.}~\bibnamefont {van
  Zuiden}}, \bibinfo {author} {\bibfnamefont {J.}~\bibnamefont {Paulose}},
  \bibinfo {author} {\bibfnamefont {W.}~\bibnamefont {Irvine}}, \bibinfo
  {author} {\bibfnamefont {D.}~\bibnamefont {Bartolo}},\ and\ \bibinfo {author}
  {\bibfnamefont {V.}~\bibnamefont {Vitelli}},\ }\bibfield  {title} {\bibinfo
  {title} {Spatiotemporal order and emergent edge currents in active spinner
  materials},\ }\href@noop {} {\bibfield  {journal} {\bibinfo  {journal} {Proc.
  Natl. Acad. Sci. USA}\ }\textbf {\bibinfo {volume} {113}},\ \bibinfo {pages}
  {12919} (\bibinfo {year} {2016})}\BibitemShut {NoStop}%
\bibitem [{\citenamefont {Shen}\ and\ \citenamefont {Lintuvuori}(2020)}]{SL20}%
  \BibitemOpen
  \bibfield  {author} {\bibinfo {author} {\bibfnamefont {Z.}~\bibnamefont
  {Shen}}\ and\ \bibinfo {author} {\bibfnamefont {J.~S.}\ \bibnamefont
  {Lintuvuori}},\ }\bibfield  {title} {\bibinfo {title} {Two-phase
  crystallization in a carpet of inertial spinners},\ }\href@noop {} {\bibfield
   {journal} {\bibinfo  {journal} {Phys. Rev. Lett.}\ }\textbf {\bibinfo
  {volume} {125}},\ \bibinfo {pages} {228002} (\bibinfo {year}
  {2020})}\BibitemShut {NoStop}%
\bibitem [{\citenamefont {Huang}\ \emph {et~al.}(2020)\citenamefont {Huang},
  \citenamefont {Menzel},\ and\ \citenamefont {L\"owen}}]{HML20}%
  \BibitemOpen
  \bibfield  {author} {\bibinfo {author} {\bibfnamefont {Z.~F.}\ \bibnamefont
  {Huang}}, \bibinfo {author} {\bibfnamefont {A.~M.}\ \bibnamefont {Menzel}},\
  and\ \bibinfo {author} {\bibfnamefont {H.}~\bibnamefont {L\"owen}},\
  }\bibfield  {title} {\bibinfo {title} {Dynamical crystallites of active
  chiral particles},\ }\href@noop {} {\bibfield  {journal} {\bibinfo  {journal}
  {Phys. Rev. Lett.}\ }\textbf {\bibinfo {volume} {125}},\ \bibinfo {pages}
  {218002} (\bibinfo {year} {2020})}\BibitemShut {NoStop}%
\bibitem [{\citenamefont {Goto}\ and\ \citenamefont {Tanaka}(2015)}]{YH15}%
  \BibitemOpen
  \bibfield  {author} {\bibinfo {author} {\bibfnamefont {Y.}~\bibnamefont
  {Goto}}\ and\ \bibinfo {author} {\bibfnamefont {H.}~\bibnamefont {Tanaka}},\
  }\bibfield  {title} {\bibinfo {title} {Purely hydrodynamic ordering of
  rotating disks at a finite reynolds number},\ }\href@noop {} {\bibfield
  {journal} {\bibinfo  {journal} {Nat. Commun.}\ }\textbf {\bibinfo {volume}
  {6}},\ \bibinfo {pages} {5994} (\bibinfo {year} {2015})}\BibitemShut
  {NoStop}%
\bibitem [{\citenamefont {Workamp}\ \emph {et~al.}(2018)\citenamefont
  {Workamp}, \citenamefont {Ramirez}, \citenamefont {Daniels},\ and\
  \citenamefont {Dijksman}}]{WRDD18}%
  \BibitemOpen
  \bibfield  {author} {\bibinfo {author} {\bibfnamefont {M.}~\bibnamefont
  {Workamp}}, \bibinfo {author} {\bibfnamefont {G.}~\bibnamefont {Ramirez}},
  \bibinfo {author} {\bibfnamefont {K.}~\bibnamefont {Daniels}},\ and\ \bibinfo
  {author} {\bibfnamefont {J.~A.}\ \bibnamefont {Dijksman}},\ }\bibfield
  {title} {\bibinfo {title} {Symmetry-reversals in chiral active matter},\
  }\href@noop {} {\bibfield  {journal} {\bibinfo  {journal} {Soft Matter}\
  }\textbf {\bibinfo {volume} {14}},\ \bibinfo {pages} {5572} (\bibinfo {year}
  {2018})}\BibitemShut {NoStop}%
\bibitem [{\citenamefont {Dasbiswas}\ \emph {et~al.}(2018)\citenamefont
  {Dasbiswas}, \citenamefont {Mandadapu},\ and\ \citenamefont
  {Vaikuntanathan}}]{DMV18}%
  \BibitemOpen
  \bibfield  {author} {\bibinfo {author} {\bibfnamefont {K.}~\bibnamefont
  {Dasbiswas}}, \bibinfo {author} {\bibfnamefont {K.}~\bibnamefont
  {Mandadapu}},\ and\ \bibinfo {author} {\bibfnamefont {S.}~\bibnamefont
  {Vaikuntanathan}},\ }\bibfield  {title} {\bibinfo {title} {Topological
  localization in out-of-equilibrium dissipative systems},\ }\href@noop {}
  {\bibfield  {journal} {\bibinfo  {journal} {Proc. Natl. Acad. Sci. USA}\
  }\textbf {\bibinfo {volume} {115}},\ \bibinfo {pages} {E9031} (\bibinfo
  {year} {2018})}\BibitemShut {NoStop}%
\bibitem [{\citenamefont {Souslov}\ \emph {et~al.}(2019)\citenamefont
  {Souslov}, \citenamefont {Dasbiswas}, \citenamefont {Fruchart}, \citenamefont
  {Vaikuntanathan},\ and\ \citenamefont {Vitelli}}]{SDFVV19}%
  \BibitemOpen
  \bibfield  {author} {\bibinfo {author} {\bibfnamefont {A.}~\bibnamefont
  {Souslov}}, \bibinfo {author} {\bibfnamefont {K.}~\bibnamefont {Dasbiswas}},
  \bibinfo {author} {\bibfnamefont {M.}~\bibnamefont {Fruchart}}, \bibinfo
  {author} {\bibfnamefont {S.}~\bibnamefont {Vaikuntanathan}},\ and\ \bibinfo
  {author} {\bibfnamefont {V.}~\bibnamefont {Vitelli}},\ }\bibfield  {title}
  {\bibinfo {title} {Topological waves in fluids with odd viscosity},\
  }\href@noop {} {\bibfield  {journal} {\bibinfo  {journal} {Phys. Rev. Lett.}\
  }\textbf {\bibinfo {volume} {122}},\ \bibinfo {pages} {128001} (\bibinfo
  {year} {2019})}\BibitemShut {NoStop}%
\bibitem [{\citenamefont {Casiulis}\ \emph {et~al.}(2020)\citenamefont
  {Casiulis}, \citenamefont {Tarzia}, \citenamefont {Cugliandolo},\ and\
  \citenamefont {Dauchot}}]{CTCD20}%
  \BibitemOpen
  \bibfield  {author} {\bibinfo {author} {\bibfnamefont {M.}~\bibnamefont
  {Casiulis}}, \bibinfo {author} {\bibfnamefont {M.}~\bibnamefont {Tarzia}},
  \bibinfo {author} {\bibfnamefont {L.~F.}\ \bibnamefont {Cugliandolo}},\ and\
  \bibinfo {author} {\bibfnamefont {O.}~\bibnamefont {Dauchot}},\ }\bibfield
  {title} {\bibinfo {title} {Velocity and speed correlations in hamiltonian
  flocks},\ }\href@noop {} {\bibfield  {journal} {\bibinfo  {journal} {Phys.
  Rev. Lett.}\ }\textbf {\bibinfo {volume} {124}},\ \bibinfo {pages} {198001}
  (\bibinfo {year} {2020})}\BibitemShut {NoStop}%
\bibitem [{\citenamefont {Monod}(1973)}]{M73}%
  \BibitemOpen
  \bibfield  {author} {\bibinfo {author} {\bibfnamefont {J.}~\bibnamefont
  {Monod}},\ }\href@noop {} {\emph {\bibinfo {title} {Le hasard et la
  n\'ecessit\'e}}}\ (\bibinfo  {publisher} {Seuil},\ \bibinfo {year}
  {1973})\BibitemShut {NoStop}%
\bibitem [{\citenamefont {Kokot}\ \emph {et~al.}(2017)\citenamefont {Kokot},
  \citenamefont {Das}, \citenamefont {Winkler}, \citenamefont {Gompper},
  \citenamefont {Aranson},\ and\ \citenamefont {Snezhko}}]{GAS17}%
  \BibitemOpen
  \bibfield  {author} {\bibinfo {author} {\bibfnamefont {G.}~\bibnamefont
  {Kokot}}, \bibinfo {author} {\bibfnamefont {S.}~\bibnamefont {Das}}, \bibinfo
  {author} {\bibfnamefont {R.}~\bibnamefont {Winkler}}, \bibinfo {author}
  {\bibfnamefont {G.}~\bibnamefont {Gompper}}, \bibinfo {author} {\bibfnamefont
  {I.}~\bibnamefont {Aranson}},\ and\ \bibinfo {author} {\bibfnamefont
  {A.}~\bibnamefont {Snezhko}},\ }\bibfield  {title} {\bibinfo {title} {Active
  turbulence in a gas of self-assembled spinners},\ }\href@noop {} {\bibfield
  {journal} {\bibinfo  {journal} {Proc. Natl. Acad. Sci. USA}\ }\textbf
  {\bibinfo {volume} {114}},\ \bibinfo {pages} {12870} (\bibinfo {year}
  {2017})}\BibitemShut {NoStop}%
\bibitem [{\citenamefont {Kokot}\ and\ \citenamefont {Snezhko}(2018)}]{KS18}%
  \BibitemOpen
  \bibfield  {author} {\bibinfo {author} {\bibfnamefont {G.}~\bibnamefont
  {Kokot}}\ and\ \bibinfo {author} {\bibfnamefont {A.}~\bibnamefont
  {Snezhko}},\ }\bibfield  {title} {\bibinfo {title} {Manipulation of emergent
  vortices in swarms of magnetic rollers},\ }\href@noop {} {\bibfield
  {journal} {\bibinfo  {journal} {Nat. Commun.}\ }\textbf {\bibinfo {volume}
  {9}},\ \bibinfo {pages} {2344} (\bibinfo {year} {2018})}\BibitemShut
  {NoStop}%
\bibitem [{\citenamefont {Bäuerle}\ \emph {et~al.}(2020)\citenamefont
  {Bäuerle}, \citenamefont {Löffler},\ and\ \citenamefont
  {Bechinger}}]{BLB20}%
  \BibitemOpen
  \bibfield  {author} {\bibinfo {author} {\bibfnamefont {T.}~\bibnamefont
  {Bäuerle}}, \bibinfo {author} {\bibfnamefont {R.~C.}\ \bibnamefont
  {Löffler}},\ and\ \bibinfo {author} {\bibfnamefont {C.}~\bibnamefont
  {Bechinger}},\ }\bibfield  {title} {\bibinfo {title} {Formation of stable and
  responsive collective states in suspensions of active colloids},\ }\href@noop
  {} {\bibfield  {journal} {\bibinfo  {journal} {Nat. Commun.}\ }\textbf
  {\bibinfo {volume} {11}},\ \bibinfo {pages} {2547} (\bibinfo {year}
  {2020})}\BibitemShut {NoStop}%
\bibitem [{\citenamefont {Liu}\ \emph {et~al.}(2020)\citenamefont {Liu},
  \citenamefont {Zhu}, \citenamefont {Zeng}, \citenamefont {Du}, \citenamefont
  {Ning}, \citenamefont {Wang}, \citenamefont {Chen}, \citenamefont {Lu},
  \citenamefont {Zheng}, \citenamefont {Ye},\ and\ \citenamefont
  {Yang}}]{ZYY20}%
  \BibitemOpen
  \bibfield  {author} {\bibinfo {author} {\bibfnamefont {P.}~\bibnamefont
  {Liu}}, \bibinfo {author} {\bibfnamefont {H.}~\bibnamefont {Zhu}}, \bibinfo
  {author} {\bibfnamefont {Y.}~\bibnamefont {Zeng}}, \bibinfo {author}
  {\bibfnamefont {G.}~\bibnamefont {Du}}, \bibinfo {author} {\bibfnamefont
  {L.}~\bibnamefont {Ning}}, \bibinfo {author} {\bibfnamefont {D.}~\bibnamefont
  {Wang}}, \bibinfo {author} {\bibfnamefont {K.}~\bibnamefont {Chen}}, \bibinfo
  {author} {\bibfnamefont {Y.}~\bibnamefont {Lu}}, \bibinfo {author}
  {\bibfnamefont {N.}~\bibnamefont {Zheng}}, \bibinfo {author} {\bibfnamefont
  {F.}~\bibnamefont {Ye}},\ and\ \bibinfo {author} {\bibfnamefont
  {M.}~\bibnamefont {Yang}},\ }\bibfield  {title} {\bibinfo {title}
  {Oscillating collective motion of active rotors in confinement},\ }\href@noop
  {} {\bibfield  {journal} {\bibinfo  {journal} {Proc. Natl. Acad. Sci. USA}\
  }\textbf {\bibinfo {volume} {117}},\ \bibinfo {pages} {11901} (\bibinfo
  {year} {2020})}\BibitemShut {NoStop}%
\bibitem [{\citenamefont {{Van Dyke}}(1982)}]{vD82}%
  \BibitemOpen
  \bibfield  {author} {\bibinfo {author} {\bibfnamefont {M.}~\bibnamefont {{Van
  Dyke}}},\ }\href@noop {} {\emph {\bibinfo {title} {An album of fluid
  motion}}}\ (\bibinfo  {publisher} {The Parabolic Press},\ \bibinfo {address}
  {Stanford, CA, USA},\ \bibinfo {year} {1982})\BibitemShut {NoStop}%
\bibitem [{\citenamefont {{Vega Reyes}}\ and\ \citenamefont
  {Urbach}(2009)}]{VU09}%
  \BibitemOpen
  \bibfield  {author} {\bibinfo {author} {\bibfnamefont {F.}~\bibnamefont
  {{Vega Reyes}}}\ and\ \bibinfo {author} {\bibfnamefont {J.~S.}\ \bibnamefont
  {Urbach}},\ }\bibfield  {title} {\bibinfo {title} {Steady base states
  navier–stokes granular hydrodynamics with boundary heating and shear},\
  }\href@noop {} {\bibfield  {journal} {\bibinfo  {journal} {J. Fluid Mech.}\
  }\textbf {\bibinfo {volume} {636}},\ \bibinfo {pages} {279} (\bibinfo {year}
  {2009})}\BibitemShut {NoStop}%
\bibitem [{\citenamefont {Chaikin}\ and\ \citenamefont
  {Lubensky}(2000)}]{CL00}%
  \BibitemOpen
  \bibfield  {author} {\bibinfo {author} {\bibfnamefont {P.~M.}\ \bibnamefont
  {Chaikin}}\ and\ \bibinfo {author} {\bibfnamefont {T.~C.}\ \bibnamefont
  {Lubensky}},\ }\href@noop {} {\emph {\bibinfo {title} {Principles of
  Condensed Matter Physics}}}\ (\bibinfo  {publisher} {Cambridge University
  Press},\ \bibinfo {address} {Cambridge, UK},\ \bibinfo {year}
  {2000})\BibitemShut {NoStop}%
\bibitem [{\citenamefont {Foerster}\ \emph {et~al.}(1994)\citenamefont
  {Foerster}, \citenamefont {Louge}, \citenamefont {Chang},\ and\ \citenamefont
  {Allia}}]{FLCA94}%
  \BibitemOpen
  \bibfield  {author} {\bibinfo {author} {\bibfnamefont {S.~F.}\ \bibnamefont
  {Foerster}}, \bibinfo {author} {\bibfnamefont {M.~Y.}\ \bibnamefont {Louge}},
  \bibinfo {author} {\bibfnamefont {H.}~\bibnamefont {Chang}},\ and\ \bibinfo
  {author} {\bibfnamefont {K.}~\bibnamefont {Allia}},\ }\bibfield  {title}
  {\bibinfo {title} {Measurements of the collision properties of small
  spheres},\ }\href@noop {} {\bibfield  {journal} {\bibinfo  {journal} {Phys.
  Fluids}\ }\textbf {\bibinfo {volume} {8}},\ \bibinfo {pages} {1108} (\bibinfo
  {year} {1994})}\BibitemShut {NoStop}%
\bibitem [{sup()}]{suppl}%
  \BibitemOpen
  \href@noop {} {}\bibinfo {note} {See Supplementary Material at [URL will be
  inserted by publisher] for more information on experimental methods,
  experiments movie clips, and additional figures}\BibitemShut {NoStop}%
\bibitem [{\citenamefont {{Vega Reyes}}\ \emph {et~al.}(2014)\citenamefont
  {{Vega Reyes}}, \citenamefont {Santos},\ and\ \citenamefont
  {Kremer}}]{VSK14}%
  \BibitemOpen
  \bibfield  {author} {\bibinfo {author} {\bibfnamefont {F.}~\bibnamefont
  {{Vega Reyes}}}, \bibinfo {author} {\bibfnamefont {A.}~\bibnamefont
  {Santos}},\ and\ \bibinfo {author} {\bibfnamefont {G.~M.}\ \bibnamefont
  {Kremer}},\ }\bibfield  {title} {\bibinfo {title} {Role of roughness on the
  hydrodynamic homogeneous base state of inelastic spheres},\ }\href@noop {}
  {\bibfield  {journal} {\bibinfo  {journal} {Phys. Rev. E}\ }\textbf {\bibinfo
  {volume} {89}},\ \bibinfo {pages} {020202(R)} (\bibinfo {year}
  {2014})}\BibitemShut {NoStop}%
\bibitem [{\citenamefont {Chapman}\ and\ \citenamefont {Cowling}(1970)}]{CC70}%
  \BibitemOpen
  \bibfield  {author} {\bibinfo {author} {\bibfnamefont {C.}~\bibnamefont
  {Chapman}}\ and\ \bibinfo {author} {\bibfnamefont {T.~G.}\ \bibnamefont
  {Cowling}},\ }\href@noop {} {\emph {\bibinfo {title} {The Mathematical Theory
  of Non-Uniform Gases}}},\ \bibinfo {edition} {3rd}\ ed.\ (\bibinfo
  {publisher} {Cambridge University Press, Cambridge},\ \bibinfo {year}
  {1970})\BibitemShut {NoStop}%
\bibitem [{\citenamefont {Slepukhin}\ and\ \citenamefont
  {Levine}(2021)}]{SL21}%
  \BibitemOpen
  \bibfield  {author} {\bibinfo {author} {\bibfnamefont {V.~M.}\ \bibnamefont
  {Slepukhin}}\ and\ \bibinfo {author} {\bibfnamefont {A.~J.}\ \bibnamefont
  {Levine}},\ }\bibfield  {title} {\bibinfo {title} {Thermal schwinger effect:
  Defect production in compressed filament bundles}} (\bibinfo {year} {2021}),\
  \bibinfo {note} {preprint in {arXiv}:2103.08832}\BibitemShut {NoStop}%
\bibitem [{\citenamefont {Sonine}(1988)}]{S88}%
  \BibitemOpen
  \bibfield  {author} {\bibinfo {author} {\bibfnamefont {N.}~\bibnamefont
  {Sonine}},\ }\bibfield  {title} {\bibinfo {title} {Two-dimensional melting},\
  }\href@noop {} {\bibfield  {journal} {\bibinfo  {journal} {Rev. Mod. Phys.}\
  }\textbf {\bibinfo {volume} {60}},\ \bibinfo {pages} {161} (\bibinfo {year}
  {1988})}\BibitemShut {NoStop}%
\bibitem [{\citenamefont {Kanatani}(1979)}]{K79}%
  \BibitemOpen
  \bibfield  {author} {\bibinfo {author} {\bibfnamefont {K.-I.}\ \bibnamefont
  {Kanatani}},\ }\bibfield  {title} {\bibinfo {title} {A micropolar continuum
  theory for the flow of granular materials},\ }\href@noop {} {\bibfield
  {journal} {\bibinfo  {journal} {Int. J. Eng. Sci.}\ }\textbf {\bibinfo
  {volume} {17}},\ \bibinfo {pages} {419} (\bibinfo {year} {1979})}\BibitemShut
  {NoStop}%
\bibitem [{\citenamefont {Evans}\ and\ \citenamefont {Morriss}(1990)}]{EM90}%
  \BibitemOpen
  \bibfield  {author} {\bibinfo {author} {\bibfnamefont {D.~J.}\ \bibnamefont
  {Evans}}\ and\ \bibinfo {author} {\bibfnamefont {G.~P.}\ \bibnamefont
  {Morriss}},\ }\href@noop {} {\emph {\bibinfo {title} {Statistical Mechanics
  of Nonequilibrium Liquids}}}\ (\bibinfo  {publisher} {Academic Press,
  London},\ \bibinfo {year} {1990})\BibitemShut {NoStop}%
\bibitem [{\citenamefont {Brady}\ and\ \citenamefont {Bossis}(1988)}]{BB88}%
  \BibitemOpen
  \bibfield  {author} {\bibinfo {author} {\bibfnamefont {J.~F.}\ \bibnamefont
  {Brady}}\ and\ \bibinfo {author} {\bibfnamefont {G.}~\bibnamefont {Bossis}},\
  }\bibfield  {title} {\bibinfo {title} {Stokesian dynamics},\ }\href@noop {}
  {\bibfield  {journal} {\bibinfo  {journal} {Ann. Rev. Fluid Mech.}\ }\textbf
  {\bibinfo {volume} {20}},\ \bibinfo {pages} {111} (\bibinfo {year}
  {1988})}\BibitemShut {NoStop}%
\bibitem [{\citenamefont {{N. G. Van Kampen}}(1997)}]{VK97}%
  \BibitemOpen
  \bibfield  {author} {\bibinfo {author} {\bibnamefont {{N. G. Van Kampen}}},\
  }\href@noop {} {\emph {\bibinfo {title} {Stochastic Processes in Physics and
  Chemistry}}},\ \bibinfo {edition} {2nd}\ ed.\ (\bibinfo  {publisher}
  {Elsevier, Amsterdam},\ \bibinfo {year} {1997})\BibitemShut {NoStop}%
\bibitem [{\citenamefont {Brey}\ \emph {et~al.}(1998)\citenamefont {Brey},
  \citenamefont {Dufty}, \citenamefont {Kim},\ and\ \citenamefont
  {Santos}}]{BDKS98}%
  \BibitemOpen
  \bibfield  {author} {\bibinfo {author} {\bibfnamefont {J.~J.}\ \bibnamefont
  {Brey}}, \bibinfo {author} {\bibfnamefont {J.~W.}\ \bibnamefont {Dufty}},
  \bibinfo {author} {\bibfnamefont {C.~S.}\ \bibnamefont {Kim}},\ and\ \bibinfo
  {author} {\bibfnamefont {A.}~\bibnamefont {Santos}},\ }\bibfield  {title}
  {\bibinfo {title} {Hydrodynamics for granular flow at low density},\
  }\href@noop {} {\bibfield  {journal} {\bibinfo  {journal} {Phys. Rev. E}\
  }\textbf {\bibinfo {volume} {58}},\ \bibinfo {pages} {4638} (\bibinfo {year}
  {1998})}\BibitemShut {NoStop}%
\bibitem [{\citenamefont {Chauviere}\ and\ \citenamefont
  {Brazzoli}(2006)}]{CB06}%
  \BibitemOpen
  \bibfield  {author} {\bibinfo {author} {\bibfnamefont {A.}~\bibnamefont
  {Chauviere}}\ and\ \bibinfo {author} {\bibfnamefont {I.}~\bibnamefont
  {Brazzoli}},\ }\bibfield  {title} {\bibinfo {title} {On the discrete kinetic
  theory for active particles. mathematical tools},\ }\href
  {https://doi.org/https://doi.org/10.1016/j.mcm.2005.10.001} {\bibfield
  {journal} {\bibinfo  {journal} {Math. Comput. Model.}\ }\textbf {\bibinfo
  {volume} {43}},\ \bibinfo {pages} {933} (\bibinfo {year} {2006})}\BibitemShut
  {NoStop}%
\bibitem [{\citenamefont {Bowick}\ \emph {et~al.}(2022)\citenamefont {Bowick},
  \citenamefont {Fakhri}, \citenamefont {Marchetti},\ and\ \citenamefont
  {Ramaswamy}}]{BFMR22}%
  \BibitemOpen
  \bibfield  {author} {\bibinfo {author} {\bibfnamefont {M.~J.}\ \bibnamefont
  {Bowick}}, \bibinfo {author} {\bibfnamefont {N.}~\bibnamefont {Fakhri}},
  \bibinfo {author} {\bibfnamefont {M.~C.}\ \bibnamefont {Marchetti}},\ and\
  \bibinfo {author} {\bibfnamefont {S.}~\bibnamefont {Ramaswamy}},\ }\bibfield
  {title} {\bibinfo {title} {Symmetry, thermodynamics, and topology in active
  matter},\ }\href {https://doi.org/10.1103/PhysRevX.12.010501} {\bibfield
  {journal} {\bibinfo  {journal} {Phys. Rev. X}\ }\textbf {\bibinfo {volume}
  {12}},\ \bibinfo {pages} {010501} (\bibinfo {year} {2022})}\BibitemShut
  {NoStop}%
\bibitem [{\citenamefont {Zhang}\ \emph {et~al.}(2022)\citenamefont {Zhang},
  \citenamefont {Yuan}, \citenamefont {Sokolov}, \citenamefont {{Olvera de la
  Cruz}},\ and\ \citenamefont {Snezhko}}]{ZYSOS22}%
  \BibitemOpen
  \bibfield  {author} {\bibinfo {author} {\bibfnamefont {B.}~\bibnamefont
  {Zhang}}, \bibinfo {author} {\bibfnamefont {H.}~\bibnamefont {Yuan}},
  \bibinfo {author} {\bibfnamefont {A.}~\bibnamefont {Sokolov}}, \bibinfo
  {author} {\bibfnamefont {M.}~\bibnamefont {{Olvera de la Cruz}}},\ and\
  \bibinfo {author} {\bibfnamefont {A.}~\bibnamefont {Snezhko}},\ }\bibfield
  {title} {\bibinfo {title} {Polar state reversal in active fluids},\ }\href
  {https://doi.org/https://doi.org/10.1038/s41567-021-01442-6} {\bibfield
  {journal} {\bibinfo  {journal} {Nat. Phys.}\ }\textbf {\bibinfo {volume}
  {18}},\ \bibinfo {pages} {154} (\bibinfo {year} {2022})}\BibitemShut
  {NoStop}%
\bibitem [{\citenamefont {Brush}(1973)}]{B73}%
  \BibitemOpen
  \bibinfo {editor} {\bibfnamefont {S.~G.}\ \bibnamefont {Brush}},\ ed.,\
  \bibinfo {title} {Kinetic theory: The {Chapman-Enskog} solution of the
  transport equation for moderately dense gases}\ (\bibinfo  {publisher}
  {Pergamon},\ \bibinfo {address} {Oxford},\ \bibinfo {year} {1973})\ Chap.\
  \bibinfo {chapter} {{D. Hilbert: Foundations of the kinetic theory of
  gases}}, pp.\ \bibinfo {pages} {89--101},\ \bibinfo {edition} {1st}\
  ed.\BibitemShut {Stop}%
\bibitem [{\citenamefont {Sonine}(1880)}]{S80}%
  \BibitemOpen
  \bibfield  {author} {\bibinfo {author} {\bibfnamefont {N.}~\bibnamefont
  {Sonine}},\ }\bibfield  {title} {\bibinfo {title} {Recherches sur les
  fonctions cylindriques et le développement des fonctions continues en
  s\'erie},\ }\href@noop {} {\bibfield  {journal} {\bibinfo  {journal} {Math
  Ann.}\ }\textbf {\bibinfo {volume} {16}},\ \bibinfo {pages} {1} (\bibinfo
  {year} {1880})}\BibitemShut {NoStop}%
\bibitem [{\citenamefont {{van Noije}}\ and\ \citenamefont
  {Ernst}(1998)}]{NE98}%
  \BibitemOpen
  \bibfield  {author} {\bibinfo {author} {\bibfnamefont {T.~P.~C.}\
  \bibnamefont {{van Noije}}}\ and\ \bibinfo {author} {\bibfnamefont {M.~H.}\
  \bibnamefont {Ernst}},\ }\bibfield  {title} {\bibinfo {title} {Velocity
  distributions in homogeneous granular fluids: the free and the heated case},\
  }\href@noop {} {\bibfield  {journal} {\bibinfo  {journal} {Granul. Matter}\
  }\textbf {\bibinfo {volume} {1}},\ \bibinfo {pages} {57} (\bibinfo {year}
  {1998})}\BibitemShut {NoStop}%
\bibitem [{\citenamefont {Olver}\ \emph {et~al.}(2010)\citenamefont {Olver},
  \citenamefont {Lozier}, \citenamefont {Boisvert},\ and\ \citenamefont
  {Clark}}]{nist}%
  \BibitemOpen
  \bibinfo {editor} {\bibfnamefont {F.~W.~J.}\ \bibnamefont {Olver}}, \bibinfo
  {editor} {\bibfnamefont {D.~W.}\ \bibnamefont {Lozier}}, \bibinfo {editor}
  {\bibfnamefont {R.~F.}\ \bibnamefont {Boisvert}},\ and\ \bibinfo {editor}
  {\bibfnamefont {C.~W.}\ \bibnamefont {Clark}},\ eds.,\ \href@noop {} {\emph
  {\bibinfo {title} {{NIST} Handbook of Mathematical Functions}}}\ (\bibinfo
  {publisher} {NIST and Cambridge University Press},\ \bibinfo {address} {New
  York, USA},\ \bibinfo {year} {2010})\BibitemShut {NoStop}%
\bibitem [{\citenamefont {Crocker}\ and\ \citenamefont
  {Grier}(1996)}]{Crocker1996}%
  \BibitemOpen
  \bibfield  {author} {\bibinfo {author} {\bibfnamefont {J.~C.}\ \bibnamefont
  {Crocker}}\ and\ \bibinfo {author} {\bibfnamefont {D.~G.}\ \bibnamefont
  {Grier}},\ }\bibfield  {title} {\bibinfo {title} {{Methods of Digital Video
  Microscopy for Colloidal Studies}},\ }\href@noop {} {\bibfield  {journal}
  {\bibinfo  {journal} {J. Colloid Interface Sci.}\ }\textbf {\bibinfo {volume}
  {179}},\ \bibinfo {pages} {298} (\bibinfo {year} {1996})}\BibitemShut
  {NoStop}%
\end{thebibliography}%
